\pdfoutput=1
\input{glyphtounicode}
\documentclass[superscriptaddress,twocolumn,amsmath,amssymb,prl]{revtex4-2}

\usepackage{amsthm}
\usepackage{float}
\usepackage{graphicx}
\usepackage{bm}
\usepackage{amsmath}
\usepackage[colorlinks=true,citecolor=blue,urlcolor=blue]{hyperref}
\usepackage{braket}
\usepackage{booktabs}
\usepackage{adjustbox}
\renewcommand{\arraystretch}{2}
\usepackage[percent]{overpic}

\usepackage{xcolor}

\def\ket#1{| #1 \rangle}
\def\bra#1{\langle #1 |}

\definecolor{blue}{rgb}{0,0,1}
\definecolor{red}{rgb}{1,0,0}
\definecolor{green}{rgb}{0,1,0}

\usepackage{amsfonts}
\usepackage{multirow}
\usepackage{setspace}
\usepackage{rotating}
\usepackage{chngcntr}

\newtheoremstyle{myremark}
  {3pt}
  {3pt}
  {\normalfont}
  {}
  {\bfseries}
  {.}
  {0.5em}
  {}
\theoremstyle{myremark}

\colorlet{revred}{red!65!black}

\def\cd{\operatorname{cd}}
\def\total{\operatorname{total}}

\newcommand{\makecell}[2][c]{\begin{tabular}{@{}#1@{}}#2\end{tabular}}
\newcommand{\panelcaption}[1]{\par\vspace{-1mm}{\footnotesize #1}\par}

\begin{document}

\title{Quantum Optical Reinforcement Learning via Spectrum-Resolved Hong-Ou-Mandel Interference}

\author{Shaojun Wu}
\thanks{These authors contributed equally to this work.}
\affiliation{Institute of  Fundamental and Frontier Sciences, University of Electronic Science and Technology of China, Chengdu, Sichuan, 611731, China}
\affiliation{Key Laboratory of Quantum Physics and Photonic Quantum Information, Ministry of Education,
  University of Electronic Science and Technology of China, Chengdu 611731, China}

\author{Jiahua Xu}
\thanks{These authors contributed equally to this work.}
\affiliation{Institute of  Fundamental and Frontier Sciences, University of Electronic Science and Technology of China, Chengdu, Sichuan, 611731, China}
\affiliation{Key Laboratory of Quantum Physics and Photonic Quantum Information, Ministry of Education,
  University of Electronic Science and Technology of China, Chengdu 611731, China}

\author{Shan Jin}
\affiliation{Institute of  Fundamental and Frontier Sciences, University of Electronic Science and Technology of China, Chengdu, Sichuan, 611731, China}
\affiliation{Key Laboratory of Quantum Physics and Photonic Quantum Information, Ministry of Education,
 University of Electronic Science and Technology of China, Chengdu 611731, China}
\affiliation{Yangtze Delta Industrial Innovation Center of Quantum Science and Technology, Suzhou 215000, China}

\author{Zhen Yang}
\affiliation{Institute of  Fundamental and Frontier Sciences, University of Electronic Science and Technology of China, Chengdu, Sichuan, 611731, China}
\affiliation{Key Laboratory of Quantum Physics and Photonic Quantum Information, Ministry of Education,
  University of Electronic Science and Technology of China, Chengdu 611731, China}
  
\author{Yifang Xu}
\affiliation{Center for Quantum Information, Institute for Interdisciplinary Information Sciences, Tsinghua University, Beijing 100084, China}

\author{Chenglong You}
\affiliation{Institute of Fundamental and Frontier Sciences, University of Electronic Science and Technology of China, Chengdu, Sichuan, 611731, China}
\affiliation{Key Laboratory of Quantum Physics and Photonic Quantum Information, Ministry of Education, University of Electronic Science and Technology of China, Chengdu 611731, China}

\author{Guangwei Deng}
\affiliation{Institute of Fundamental and Frontier Sciences, University of Electronic Science and Technology of China, Chengdu, Sichuan, 611731, China}
\affiliation{Key Laboratory of Quantum Physics and Photonic Quantum Information, Ministry of Education, University of Electronic Science and Technology of China, Chengdu 611731, China}

\author{Luyan Sun}
\email{luyansun@tsinghua.edu.cn}
\affiliation{Center for Quantum Information, Institute for Interdisciplinary Information Sciences, Tsinghua University, Beijing 100084, China}
\affiliation{Hefei National Laboratory, Hefei 230088, China}

\author{Chang-Ling Zou}
\email{clzou321@ustc.edu.cn}
\affiliation{Laboratory of Quantum Information, University of Science and Technology of China, Hefei 230026, China}
\affiliation{Hefei National Laboratory, Hefei 230088, China}

\author{Xiaoting Wang}
\email{xiaoting@uestc.edu.cn}
\affiliation{Institute of Fundamental and Frontier Sciences, University of Electronic Science and Technology of China, Chengdu, Sichuan, 611731, China}
\affiliation{Key Laboratory of Quantum Physics and Photonic Quantum Information, Ministry of Education,
 University of Electronic Science and Technology of China, Chengdu 611731, China}

\date{\today}

\begin{abstract}
Hong-Ou-Mandel (HOM) interference-based optical neural networks can offer complexity advantages on benchmark learning tasks, but conventional readout compresses the coincidence spectrum into a single scalar, limiting its use in complex settings such as continuous-action reinforcement learning. Here we introduce a spectrum-resolved HOM (SR-HOM) architecture that promotes the photons' spectral degrees of freedom to a trainable computational resource and use it to construct a compact optical actor-critic agent. Diagonal spectral responses generate continuous actions, while higher-order spectral correlations provide nonlinear state-action features for value estimation. Across five continuous-control benchmarks, SR-HOM outperforms parameter-matched multilayer-perceptron baselines, including a \(4.4\times\) improvement in sample efficiency and a \(74.0\%\) increase in best 100-episode moving-average return for LunarLanderContinuous-v3. Applied to online calibration of drifted tunable-coupler CZ and iSWAP gates for transmon qubits, simulations show it restores fidelities to \(0.9917\) and \(0.9952\) respectively, exceeding \(99.8\%\) of their drift-free calibrated values.
\end{abstract}

\maketitle

\paragraph{Introduction.} Quantum optical platforms are natural substrates for high-dimensional information processing, since optical fields carry continuous spectral and temporal degrees of freedom and can be manipulated interferometrically with low loss and high stability \cite{Shastri2021,McMahon2023,Lin2018,Carolan2015,Slussarenko2019,Wetzstein2020,Miscuglio2020,Shen2017}. Recent advances in HOM optical neural networks implement learning directly through coincidence estimation, avoiding image reconstruction and resolution-dependent digital post-processing \cite{Bowie2024,Roncallo2025QOptClassifier,Roncallo2025QOSN,Minati2026QON,Roncallo2026Divide}. However, standard HOM readout integrates over all detected frequencies and returns only a scalar coincidence probability or visibility. Although this scalar nonlinear overlap is effective for binary discrimination and related low-dimensional tasks \cite{Hong1987,Bouchard2021,Legero2004}, it compresses the spectral information carried by the interfering modes, limiting representational capacity and making extensions to more complex tasks reliant on additional HOM units \cite{Roncallo2025QOptClassifier,Roncallo2025QOSN,Roncallo2026Divide}. This limitation becomes particularly restrictive in more demanding learning settings that require structured outputs, such as reinforcement learning (RL) with continuous action spaces~\cite{Wu2025quantum,Wu2023QSGDRL,niu2019universal,duan2016benchmarking,Sivak2022ModelFree},
where actor-critic methods rely on function approximators that map observations to continuous actions and state-action values \cite{sutton2018reinforcement,silver2014deterministic,lillicrap2016continuous,Fujimoto2018,Haarnoja2018}. A quantum optical RL implementation must therefore generate structured continuous outputs and nonlinear features for value estimation, not merely a single similarity score.

To address this limitation, we introduce a RL architecture based on SR-HOM interference. Frequency-resolved coincidence detection promotes the HOM output from a scalar visibility to a structured interference tensor, retaining spectral information that is discarded  in integrated HOM measurements \cite{Brecht2015,Lukens2017,Raymer2020,Jin2015}. The environment state is encoded in the spectral mode of a single photon and interferes with a trainable probe photon at a balanced beam splitter. Grouped diagonal components of the coincidence tensor generate continuous actions, while higher-order spectral correlations provide nonlinear state-action features for value estimation. Under ideal interference, estimating the action-relevant spectral marginals to accuracy $\varepsilon$ with confidence $1-\delta$ requires $O(\varepsilon^{-2}\log k)$ coincidence samples for spectral resolution $k$, independent of the ambient state dimension. Numerical experiments on representative continuous-control benchmarks show stable training in a strongly capacity-limited regime: parameter-matched MLP actor-critic baselines fail to converge reliably, whereas the SR-HOM architecture remains trainable and achieves faster, more stable learning. 

As a physically motivated application, we further deploy the same SR-HOM architecture for online calibration of tunable-coupler two-qubit gates. In superconducting transmon processors, flux-driven tunable couplers modulate the interaction between qubits and thereby realize entangling gates~\cite{Yan2018,Koch2007}. Slow drift in device parameters and distortions in the control line can progressively reduce gate fidelity and increase leakage, eventually requiring full recalibration. Such procedures interrupt processor operation and can incur substantial experimental overhead, particularly as system size and calibration complexity increase~\cite{Sung2021TunableCoupler,Li2025PulseCalibration,AutoCalib112Qubit,ScalingCalibrationReport}. We therefore introduce a RL correction protocol designed to extend the interval between full recalibrations. Using only measurement-derived gate diagnostics, the SR-HOM agent incrementally updates the control pulse during processor idle periods or between computational workloads, without direct access to the underlying drift parameters or gate fidelity. Simulations show that in our scheme, the learned policy tracks continuously evolving parameter drift and pulse distortion with low experimental overhead, restoring the CZ and iSWAP fidelities close to their initially calibrated reference values while simultaneously reducing leakage.

\paragraph{Standard HOM quantum optical neural network.} Recent work has shown that HOM interference can realize a quantum optical neuron by encoding the input and trainable parameters into single-photon modes that interfere on a balanced beam splitter \cite{Roncallo2025QOptClassifier,Mandel1995,Ou2007}. For an input wavepacket $\ket{1_{\psi_{x}}}$ and trainable probe $\ket{1_{\phi_{\lambda}}}$, standard HOM detection integrates all frequencies and gives a single coincidence probability
\begin{equation}
P_{\mathrm{c}}(\bm{x},\bm{\lambda}) =\frac{1}{2}\left(1-\left|\int d\omega\,\psi^*_{\bm{x}}(\omega)\phi_{\bm{\lambda}}(\omega)\right|^2\right).
\label{eq:standard_hom_pc}
\end{equation}
The output is a nonlinear function of the optical-mode overlap. With multiple probes, this implements a shallow quantum optical network whose inference is performed directly by coincidence estimation, without reconstructing the input field \cite{Roncallo2025QOptClassifier,Roncallo2025QOSN}.

By enabling inference directly from coincidence statistics without reconstructing the input states, this framework underlies the favorable complexity scaling emphasized in HOM-based optical learning proposals \cite{Roncallo2025QOptClassifier,Minati2026QON,Roncallo2026Divide}. Its limitation is that all spectral information is integrated into one coincidence statistic. The standard HOM neuron is therefore expressive enough for overlap-based discrimination but poorly suited to tasks requiring structured continuous outputs or rich nonlinear features from high-dimensional inputs.

\begin{figure*}[t]
    \centering
    \includegraphics[width=0.8\textwidth]{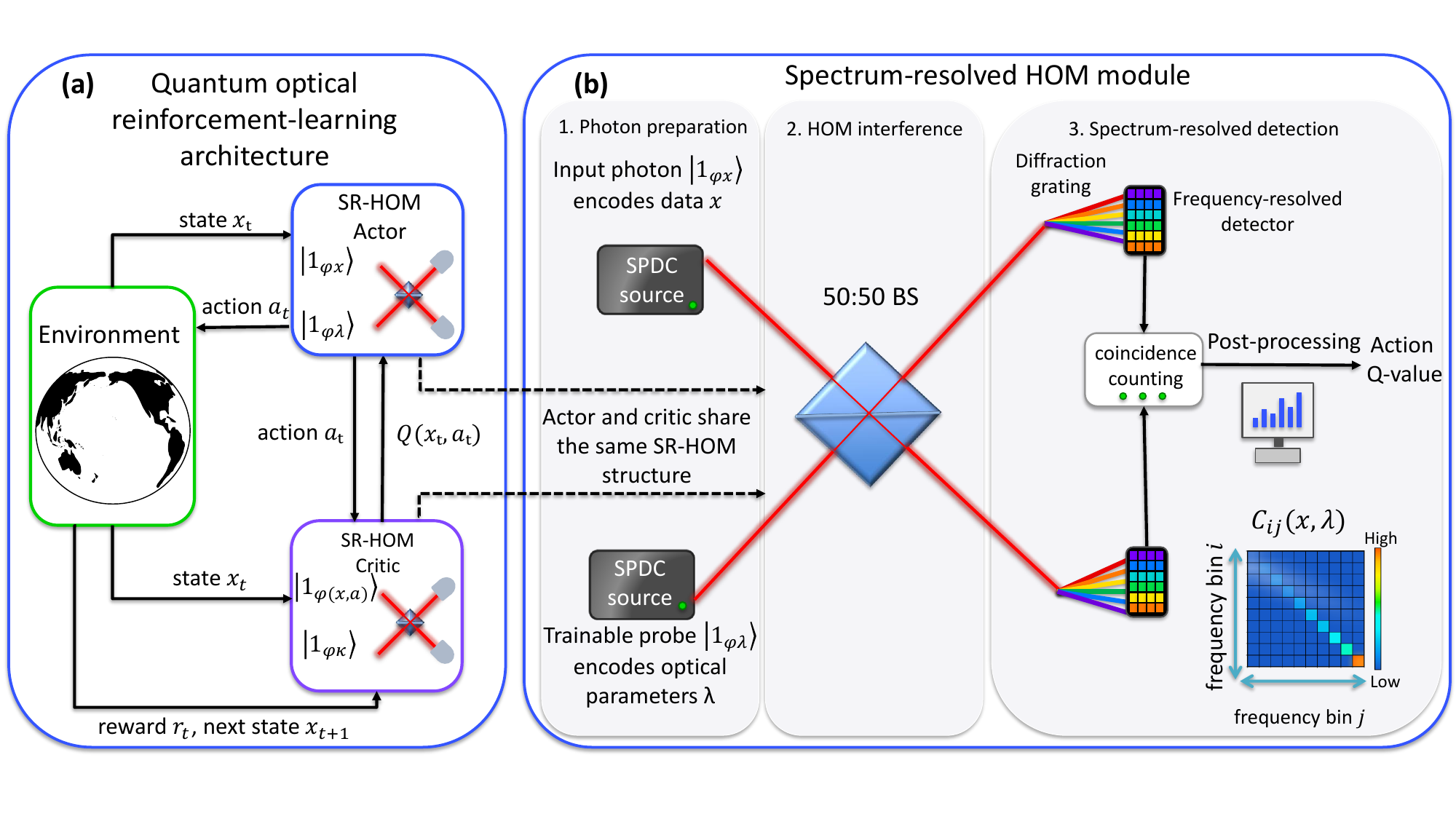}
    \caption{
    Quantum optical reinforcement learning with SR-HOM modules.
    (a) Actor-critic architecture. The environment provides the state $x_t$ to the SR-HOM actor, which outputs the continuous action $a_t$; the SR-HOM critic evaluates the state-action pair and returns $Q(x_t,a_t)$.
    (b) SR-HOM module. A state-encoded input photon and a trainable probe photon interfere at a balanced $50{:}50$ beam splitter, followed by frequency-resolved coincidence detection to produce the tensor $C_{ij}(x,\lambda)$. Grouped diagonal entries are used for action generation, while the tensor response supports value evaluation.
    }
    \label{fig:sr_hom_architecture}
\end{figure*}

\paragraph{Spectrum-resolved HOM quantum optical neural network.} To overcome this limitation, we introduce a spectrum-resolved HOM quantum optical neural network that retains the frequency information averaged out in standard HOM detection. The basic interferometer is unchanged: the input spectral mode $\psi_{\bm{x}}(\omega)$ and trainable probe $\phi_{\bm{\lambda}}(\omega)$ interfere on a balanced beam splitter, with classical data and optical weights encoded in an orthonormal spectral basis~\cite{Jin2015,Gerrits2015PRA,YepizGraciano2020PR}. Instead of integrating over all frequencies, we consider the frequency-resolved coincidence density
\begin{equation}
P_{cd}(\omega_1,\omega_2;\bm{x},\bm{\lambda})
=
\frac{1}{4}
\left|
\psi_{\bm{x}}(\omega_1)\phi_{\bm{\lambda}}(\omega_2)
-
\psi_{\bm{x}}(\omega_2)\phi_{\bm{\lambda}}(\omega_1)
\right|^2.
\label{eq:srhom_density}
\end{equation}
This quantity retains the detailed spectral structure of the two-photon interference. Because it arises from the squared two-photon amplitude, it depends nonlinearly on both the input and probe mode coefficients and therefore provides a much richer optical feature map than the single scalar coincidence rate of Eq.~(\ref{eq:standard_hom_pc}).

To obtain a finite-dimensional output, we partition the detection bandwidth into $k$ disjoint frequency bins and define the bin-resolved coincidence tensor
\begin{equation}
C_{ij}(\bm{x},\bm{\lambda})
=
\int_{\Omega_i} d\omega_1
\int_{\Omega_j} d\omega_2\,
P_{cd}(\omega_1,\omega_2;\bm{x},\bm{\lambda}),
\label{eq:coincidence_tensor}
\end{equation}
normalized such that $\sum_{i,j} C_{ij}=1$. The tensor $C(\bm{x},\bm{\lambda})\in\mathbb{R}^{k\times k}$ is therefore a structured interference representation generated directly by two-photon exchange interference. The standard HOM neuron is recovered by coarse-graining this tensor to one integrated coincidence probability. In contrast, SR-HOM provides a tensor-valued photonic representation: diagonal components can generate smooth continuous outputs, while off-diagonal and higher-order spectral correlations supply nonlinear features for value estimation. Thus the architecture preserves the compact HOM interference backbone while increasing the representational capacity needed for continuous-control reinforcement learning.

\paragraph{Quantum optical reinforcement learning.} We use a deterministic actor-critic architecture in which both function approximators are SR-HOM modules, as shown in Fig.~\ref{fig:sr_hom_architecture}(a). Each module receives spectrally encoded input photons, which interfere with independent trainable probes at a balanced beam splitter, and reads out the bin-resolved tensor $C_{ij}(x,\lambda)$. The actor uses an SR-HOM module to construct the continuous action from the diagonal entries of this tensor. Specifically, we define
\begin{equation}
p_i(\bm{x},\bm{\lambda}) = C_{ii}(\bm{x},\bm{\lambda}),
\end{equation}
and partition the index set into disjoint groups $\{\mathcal{G}_r\}_{r=1}^{m}$, from which the $m$-dimensional action is obtained as
\begin{equation}
\bm{a_r}(\bm{x},\bm{\lambda})
=
g_{r}\!\left(
\sum_{i\in\mathcal{G}_r} p_i(\bm{x},\bm{\lambda})
\right),
\qquad
r=1,\ldots,m,
\label{eq:actor_action}
\end{equation}
where $g$ is a fixed monotone map to the target action range. The action dimensionality is set by the spectral grouping, while the response is governed by the structured interference encoded in the tensor rather than by a scalar overlap. The actor can thus produce structured continuous outputs directly from spectrum-resolved interference statistics. The critic, on the other hand, uses a second SR-HOM module with a joint encoding of the state-action pair $\bm x= (\bm{s}, \bm{a})$ to estimate the expected return obtained by taking action $\bm{a}$ in state $\bm{s}$ and subsequently following
the current policy. Its spectrum-resolved readout is the
tensor
\begin{equation}
q_i(\bm{x}, \bm{\kappa})
=
C_{ii}(\bm{x}, \bm{\kappa}),
\end{equation}
where $\bm{\kappa}$ denotes the trainable parameters of the SR-HOM
module. The resulting tensor is mapped to a scalar action-value estimate through a trainable linear readout,
\begin{equation}
Q(\bm{x}, \bm{\kappa})
=
f_{\bm{W},b}
\left(
\sum_{i} q_i(\bm{x}, \bm{\kappa})
\right),
\label{eq:q_abstract}
\end{equation}
where $f_{\bm{W},b}$ is parameterized by trainable weights $\bm{W}$ and
bias $b$. The actor and critic therefore use the same physical primitive with complementary readouts: grouped diagonal entries map states to actions, and the second module maps state-action pairs to scalars through the tensor response.

As shown in Fig.~\ref{fig:sr_hom_architecture}(b), realizing an SR-HOM module calls for three capabilities, each individually demonstrated in photonic experiments: (i) encoding the environment state into the spectral wavefunction $\psi_x(\omega)$ of a single photon; (ii) a reconfigurable trainable probe $\phi_\lambda(\omega)$ whose spectral amplitude and phase embody the learnable parameters $\lambda$; and (iii) frequency-resolved coincidence detection of the tensor $C_{ij}$. 
The programmable spectral phase and amplitude needed to imprint the data $x$ onto the single-photon spectral wavefunction can be realized via electro-optic time lenses and Fourier-domain pulse shaping~\cite{Lavoie2013,Karpinski2017,Sosnicki2023}, and arbitrary unitaries on frequency-bin-encoded photons have been demonstrated using photonic quantum processors~\cite{Lukens2017,Lu23}. Frequency-resolved detection that promotes the scalar HOM dip to the tensor $C_{ij}$ is provided by dispersive grating-based spectral demultiplexing onto a single-photon-detector array. Notably, the module can be realized in an integrated platform, e.g., thin-film lithium niobate photonic chips~\cite{Buddhiraju2021,Zhu2022}, where an implementation of a universal frequency-encoded gate set on chip has been reported recently~\cite{Yang2026}.

\paragraph{Complexity Analysis.} Each HOM trial either produces no coincidence or, upon coincidence, returns a detected bin pair $(i,j)\in\{1,\ldots,k\}^2$. Conditioned on coincidence events, these bin pairs follow a categorical distribution with probabilities $C_{ij}(\boldsymbol{x},\bm\lambda)$. Since the action construction depends only on the diagonal entries $p_i(\boldsymbol{x},\bm\lambda)=C_{ii}(\boldsymbol{x},\bm\lambda)$, we define $Z_i^{(t)}=\mathbf{1}[(i_t,j_t)=(i,i)]$ for the $t$-th accepted coincidence sample. After $M$ independent coincidence samples, the empirical estimator is $\widehat p_i=M^{-1}\sum_{t=1}^M Z_i^{(t)}$. By Hoeffding's inequality ~\cite{Hoeffding1963},
\[
\Pr\!\left(|\widehat p_i-p_i|\ge\varepsilon\right)\le 2e^{-2M\varepsilon^2}.
\]
Applying a union bound over the $k$ diagonal entries yields
\begin{equation}
M \;\ge\; \frac{1}{2\varepsilon^2}\log\frac{2k}{\delta},
\label{eq:freq_sample_bound_final}
\end{equation}
which guarantees $\max_{i} |\widehat p_i-p_i|\le \varepsilon$ simultaneously for all diagonal bins with probability at least $1-\delta$.
\begin{figure*}[t]
    \centering
    \includegraphics[width=0.93\linewidth]{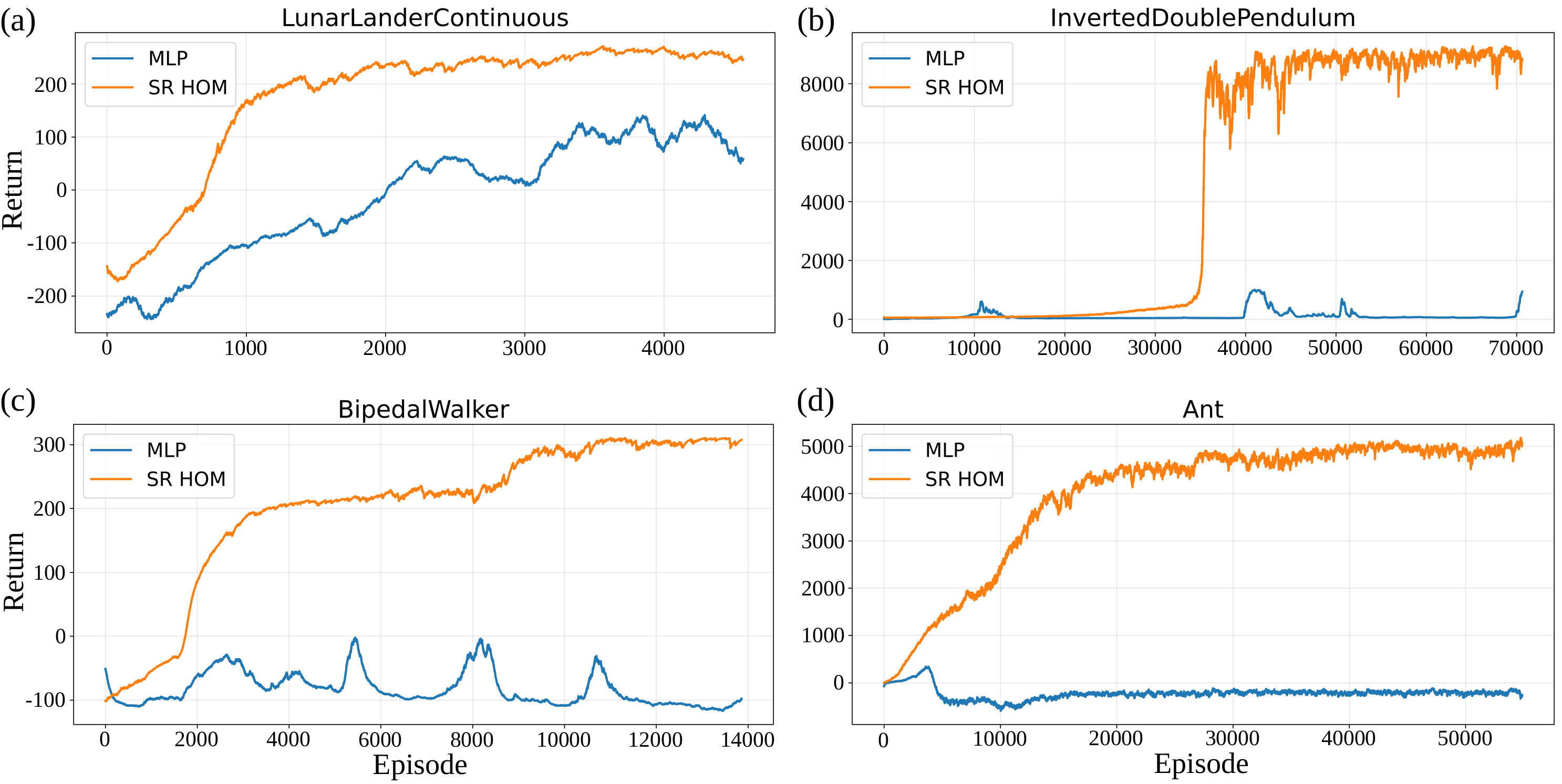}
    \caption{
    Smoothed training-return comparison between the parameter-matched MLP baseline and the proposed SR-HOM agent on four continuous-control benchmarks:
    (a) LunarLanderContinuous-v3,
    (b) InvertedDoublePendulum-v5,
    (c) BipedalWalker-v3,
    and (d) Ant-v5.
    Across all environments, the SR-HOM agent learns more effectively and reaches substantially better final performance.
    Panels (b), (c), and (d) show that the advantage of SR-HOM agent becomes more pronounced on more challenging tasks.
    }
    \label{fig:sim_compare}
\end{figure*}
The action components are sums over diagonal groups followed by the fixed monotone rescaling $g$.
Hence, if the group sizes are bounded and $g$ is Lipschitz, Eq.~(\ref{eq:freq_sample_bound_final}) implies $O(\varepsilon)$ accuracy for the action vector up to a constant factor.

If the coincidence probability is $P_c(\boldsymbol{x},\bm\lambda)$, the expected number of consumed photon pairs is
\begin{equation}
N_{\mathrm{pairs}}
=
O\!\left(
\frac{1}{P_c(\boldsymbol{x},\bm\lambda)}
\,\varepsilon^{-2}
\log\frac{k}{\delta}
\right).
\end{equation}
Thus, the action-estimation cost scales as $O(\varepsilon^{-2}\log k)$ in spectral resolution and does not scale explicitly with the state-encoding dimension $d$, since the HOM device processes the encoded optical mode directly.


\paragraph{Benchmarking and comparison.} We evaluate the SR-HOM agent on five continuous-control benchmarks from Gymnasium~\cite{brockman2016openai}, spanning low-dimensional thrust control to high-dimensional locomotion. For each task, we compare it with a parameter-matched MLP using the same reinforcement-learning algorithm. The MLP configuration is selected by hyperparameter search and multiple-seed evaluation, and we report its best run according to the best 100-episode moving-average return, thereby favoring the classical baseline. Full environment specifications, baseline-selection details, and additional learning curves are provided in the Supplemental Material.

On LunarLanderContinuous-v3, SR-HOM reaches the return threshold of \(200\) after \(666\) episodes, compared with \(2935\) episodes for the MLP, corresponding to a \(77.3\%\) reduction in episodes-to-threshold and a \(4.4\times\) improvement in sample efficiency. It also reduces the post-threshold collapse rate by \(82.3\%\), lowers curve volatility by \(13.2\%\), and improves the best 100-episode moving-average return by \(74.0\%\). Across the remaining benchmarks, SR-HOM consistently reaches higher-return regimes more rapidly and maintains more stable learning trajectories. Moreover, action generation requires only grouped diagonal entries of the coincidence tensor rather than reconstruction of the full tensor. Together with the diagonal-readout sampling bound, these results support SR-HOM as a compact yet expressive optical function approximator for continuous-action reinforcement learning.

\paragraph{Online RL calibration of tunable-coupler two-qubit gates.} We apply the SR-HOM architecture to a physically grounded continuous-control problem: online correction of tunable-coupler CZ and iSWAP gates for transmon qubits. The RL environment is a simulation of a tunable coupler system incorporating stochastic parameter drift and control-line distortion. At each calibration step, the agent receives finite-shot, gate-specific diagnostics of coherent control error and leakage and outputs bounded incremental corrections to five coupler-pulse parameters ~\cite{Li2025PulseCalibration} and two virtual-\(Z\) phases ~\cite{McKay2017EfficientZ}. The reward depends only on these measurement-derived diagnostics and action regularization; the true gate fidelity is excluded from both the observation and reward and is evaluated only for validation.

When tested on independently generated drift conditions not encountered during training, the uncorrected mean CZ and iSWAP fidelities decrease to \(0.9547\) and \(0.9640\), respectively. The RL corrections raise them to \(0.9917\) and \(0.9952\), with both exceeding \(99.8\%\) of their corresponding pulse references obtained under drift-free conditions. The mean leakage is simultaneously reduced by \(27.4\%\) for CZ and \(35.4\%\) for iSWAP. These results show that the agent can compensate drift-induced coherent control errors without direct access to either the hidden drift parameters or the gate fidelity. Further details of the physical model, diagnostic protocol, and training procedure are provided in the Supplemental Material.

\paragraph{Concluding discussion.} Taken together, our results show how photonic interference can function not only as a similarity measurement, but also as a trainable physical feature map embedded within a computational architecture~\cite{Schuld2019}. Spectrum-resolved detection promotes the conventional scalar HOM response to a tensor of nonlinear spectral correlations, while trainable spectral probes convert these correlations into task-dependent features for policy generation and value estimation. The same architecture can therefore support both standard continuous-control benchmarks and physically grounded online-calibration tasks. In this hybrid optical-digital framework, optical interference performs structured feature generation, whereas digital processing maps the measured correlations to task-specific outputs and updates the trainable parameters. In addition, the underlying principle of this work extends beyond RL. By assigning distinct computational roles to diagonal, grouped, and off-diagonal components of the SR-HOM coincidence tensor, the framework can be adapted to classification, prediction, and other decision-making tasks. Compact digital readout layers can then transform the resulting optical correlations into task-specific predictions, decisions, or control signals.

\begin{acknowledgments}
This work was supported by the National Key R\&D Program of China (Grant No.~2022YFA1405900), the National Natural Science Foundation of China (Grants No.~92265208; U2441217) and the Sichuan Science and Technology Program (Grant No.~2025YFHZ0336; 2024YFHZ0372).
\end{acknowledgments}

\bibliographystyle{apsrev4-2}
\bibliography{ref_for_combined}

@article{Shastri2021,
  author        = {Shastri, Bhavin J. and Tait, Alexander N. and Ferreira de Lima, Tiago and Pernice, Wolfram H. P. and Bhaskaran, Harish and Wright, Christopher D. and Prucnal, Paul R.},
  title         = {Photonics for artificial intelligence and neuromorphic computing},
  journal       = {Nature Photonics},
  year          = {2021},
  volume        = {15},
  pages         = {102--114},
  doi           = {10.1038/s41566-020-00754-y},
  url           = {https://www.nature.com/articles/s41566-020-00754-y}
}

@article{McMahon2023,
  author        = {Peter L. McMahon},
  title         = {The physics of optical computing},
  journal       = {Nature Reviews Physics},
  year          = {2023},
  volume        = {5},
  number        = {12},
  pages         = {717--734},
  doi           = {10.1038/s42254-023-00645-5}
}

@article{Lin2018,
  author        = {Xing Lin and Yair Rivenson and Nezih T. Yardimci and Muhammed Veli and Yi Luo and Mona Jarrahi and Aydogan Ozcan},
  title         = {All-optical machine learning using diffractive deep neural networks},
  journal       = {Science},
  year          = {2018},
  volume        = {361},
  number        = {6406},
  pages         = {1004--1008},
  doi           = {10.1126/science.aat8084}
}

@article{Carolan2015,
  author        = {Jacques Carolan and Christopher Harrold and Chris Sparrow and others},
  title         = {Universal Linear Optics on a Photonic Chip},
  journal       = {Science},
  year          = {2015},
  volume        = {349},
  number        = {6249},
  pages         = {711--716},
  url           = {https://www.science.org/doi/10.1126/science.aab3642}
}

@article{Slussarenko2019,
  author        = {Sergei Slussarenko and Geoff J. Pryde},
  title         = {Photonic Quantum Information: A Review},
  journal       = {Applied Physics Reviews},
  year          = {2019},
  volume        = {6},
  number        = {4},
  pages         = {041303},
  url           = {https://aip.scitation.org/doi/10.1063/1.5115814}
}

@article{Wetzstein2020,
  author        = {Wetzstein, Gordon and Ozcan, Aydogan and Gigan, Sylvain and Fan, Shanhui and Englund, Dirk and Solja{\v c}i{\'c}, Marin and Denz, Cornelia and Miller, David A. B. and Psaltis, Demetri},
  title         = {Inference in artificial intelligence with deep optics and photonics},
  journal       = {Nature},
  year          = {2020},
  volume        = {588},
  number        = {7836},
  pages         = {39--47},
  doi           = {10.1038/s41586-020-2973-6},
  url           = {https://www.nature.com/articles/s41586-020-2973-6}
}

@article{Miscuglio2020,
  author        = {Miscuglio, Mario and Sorger, Volker J.},
  title         = {Photonic tensor cores for machine learning},
  journal       = {Applied Physics Reviews},
  year          = {2020},
  volume        = {7},
  number        = {3},
  pages         = {031404},
  doi           = {10.1063/5.0001942},
  url           = {https://pubs.aip.org/aip/apr/article/7/3/031404/998338/Photonic-tensor-cores-for-machine-learning}
}

@article{Shen2017,
  author        = {Shen, Yichen and Harris, Nicholas C. and Skirlo, Scott and Prabhu, Mihika and Baehr-Jones, Tom and Hochberg, Michael and Sun, Xin and Zhao, Shijie and Larochelle, Hugo and Englund, Dirk and Solja{\v c}i{\'c}, Marin},
  title         = {Deep learning with coherent nanophotonic circuits},
  journal       = {Nature Photonics},
  year          = {2017},
  volume        = {11},
  pages         = {441--446},
  doi           = {10.1038/nphoton.2017.93},
  url           = {https://www.nature.com/articles/nphoton.2017.93}
}

@article{Bowie2024,
  author        = {Cassandra Bowie and Sally Shrapnel and Michael J. Kewming},
  title         = {Quantum kernel evaluation via Hong--Ou--Mandel interference},
  journal       = {Quantum Science and Technology},
  year          = {2024},
  volume        = {9},
  number        = {1},
  pages         = {015001},
  doi           = {10.1088/2058-9565/acfba9}
}

@article{Roncallo2025QOptClassifier,
  author        = {Simone Roncallo and Angela Rosy Morgillo and Chiara Macchiavello and Lorenzo Maccone and Seth Lloyd},
  title         = {Quantum optical classifier with superexponential speedup},
  journal       = {Communications Physics},
  year          = {2025},
  volume        = {8},
  pages         = {147},
  doi           = {10.1038/s42005-025-02020-5}
}

@article{Roncallo2025QOSN,
  author        = {Simone Roncallo and Angela Rosy Morgillo and Seth Lloyd and Chiara Macchiavello and Lorenzo Maccone},
  title         = {Quantum optical shallow networks},
  journal       = {arXiv preprint arXiv:2507.21036},
  year          = {2025},
  eprint        = {2507.21036},
  archivePrefix = {arXiv},
  primaryClass  = {quant-ph}
}

@article{Minati2026QON,
  author        = {Giorgio Minati and Simone Roncallo and Simone Scrofana and Angela Rosy Morgillo and Nicol{\'o} Spagnolo and Chiara Macchiavello and Lorenzo Maccone and Valeria Cimini and Fabio Sciarrino},
  title         = {Quantum Optical Neuron for Image Classification via Multiphoton Interference},
  journal       = {arXiv preprint arXiv:2603.28879},
  year          = {2026},
  eprint        = {2603.28879},
  archivePrefix = {arXiv},
  primaryClass  = {quant-ph}
}

@article{Roncallo2026Divide,
  author        = {Simone Roncallo and Angela Rosy Morgillo and Seth Lloyd and Chiara Macchiavello and Lorenzo Maccone},
  title         = {Divide et impera: hybrid multinomial classifiers from quantum binary models},
  journal       = {arXiv preprint arXiv:2604.08094},
  year          = {2026},
  eprint        = {2604.08094},
  archivePrefix = {arXiv},
  primaryClass  = {quant-ph}
}

@article{Hong1987,
  author        = {Hong, C. K. and Ou, Z. Y. and Mandel, L.},
  title         = {Measurement of subpicosecond time intervals between two photons by interference},
  journal       = {Physical Review Letters},
  year          = {1987},
  volume        = {59},
  number        = {18},
  pages         = {2044--2046},
  doi           = {10.1103/PhysRevLett.59.2044},
  url           = {https://link.aps.org/doi/10.1103/PhysRevLett.59.2044}
}

@article{Bouchard2021,
  author        = {Fr{\'e}d{\'e}ric Bouchard and Alicia Sit and Yingwen Zhang and Robert Fickler and Filippo M. Miatto and Yuan Yao and Fabio Sciarrino and Ebrahim Karimi},
  title         = {Two-photon interference: the Hong--Ou--Mandel effect},
  journal       = {Reports on Progress in Physics},
  year          = {2021},
  volume        = {84},
  number        = {1},
  pages         = {012402},
  doi           = {10.1088/1361-6633/abcd7a}
}

@article{Legero2004,
  author        = {Thomas Legero and Tatjana Wilk and Markus Hennrich and Gerhard Rempe and Axel Kuhn},
  title         = {Quantum beat of two single photons},
  journal       = {Physical Review Letters},
  year          = {2004},
  volume        = {93},
  number        = {7},
  pages         = {070503},
  doi           = {10.1103/PhysRevLett.93.070503}
}

@article{Wu2025quantum,
  author        = {Wu, Shaojun and Jin, Shan and Wen, Dingding and Han, Donghong and Wang, Xiaoting},
  title         = {Quantum reinforcement learning in continuous action space},
  journal       = {{Quantum}},
  year          = {2025},
  volume        = {9},
  pages         = {1660},
  doi           = {10.22331/q-2025-03-12-1660},
  url           = {https://doi.org/10.22331/q-2025-03-12-1660}
}

@inproceedings{Wu2023QSGDRL,
  author        = {Wu, Shaojun and Jin, Shan and Wang, Xiaoting},
  title         = {Quantum State Generation Via Deep Reinforcement Learning},
  booktitle     = {2023 IEEE International Conference on Systems, Man, and Cybernetics (SMC)},
  year          = {2023},
  pages         = {390--395},
  doi           = {10.1109/SMC53992.2023.10394265},
  url           = {https://doi.org/10.1109/SMC53992.2023.10394265}
}

@article{niu2019universal,
  author        = {Niu, Murphy Yuezhen and Boixo, Sergio and Smelyanskiy, Vadim N and Neven, Hartmut},
  title         = {Universal quantum control through deep reinforcement learning},
  journal       = {npj Quantum Information},
  year          = {2019},
  volume        = {5},
  number        = {33},
  pages         = {1--8},
  doi           = {10.1038/s41534-019-0141-3},
  url           = {https://doi.org/10.1038/s41534-019-0141-3}
}

@inproceedings{duan2016benchmarking,
  author        = {Duan, Yan and Chen, Xi and Houthooft, Rein and Schulman, John and Abbeel, Pieter},
  title         = {Benchmarking deep reinforcement learning for continuous control},
  booktitle     = {Proceedings of the 33rd International Conference on Machine Learning - Volume 48},
  year          = {2016},
  series        = {ICML'16},
  pages         = {1329--1338},
  address       = {New York, NY, USA},
  doi           = {10.5555/3045390.3045531},
  url           = {https://dl.acm.org/doi/10.5555/3045390.3045531}
}

@article{Sivak2022ModelFree,
  author        = {Sivak, V. V. and Eickbusch, A. and Liu, H. and Royer, B. and Tsioutsios, I. and Devoret, M. H.},
  title         = {Model-Free Quantum Control with Reinforcement Learning},
  journal       = {Phys. Rev. X},
  year          = {2022},
  volume        = {12},
  number        = {1},
  pages         = {011059},
  doi           = {10.1103/PhysRevX.12.011059},
  url           = {https://link.aps.org/doi/10.1103/PhysRevX.12.011059}
}

@book{sutton2018reinforcement,
  author        = {Sutton, Richard S. and Barto, Andrew G.},
  title         = {{Reinforcement Learning: An Introduction}},
  publisher     = {The MIT Press},
  year          = {2018},
  edition       = {Second},
  url           = {http://incompleteideas.net/book/the-book-2nd.html}
}

@inproceedings{silver2014deterministic,
  author        = {Silver, David and Lever, Guy and Heess, Nicolas and Degris, Thomas and Wierstra, Daan and Riedmiller, Martin},
  title         = {Deterministic Policy Gradient Algorithms},
  booktitle     = {Proceedings of the 31st International Conference on Machine Learning},
  publisher     = {PMLR},
  year          = {2014},
  volume        = {32},
  series        = {Proceedings of Machine Learning Research},
  pages         = {387--395},
  address       = {Beijing, China},
  url           = {https://proceedings.mlr.press/v32/silver14.html}
}

@misc{lillicrap2016continuous,
  author        = {Timothy P. Lillicrap and Jonathan J. Hunt and Alexander Pritzel and Nicolas Heess and Tom Erez and Yuval Tassa and David Silver and Daan Wierstra},
  title         = {{Continuous control with deep reinforcement learning}},
  year          = {2015},
  doi           = {10.48550/ARXIV.1509.02971},
  url           = {https://arxiv.org/abs/1509.02971},
  eprint        = {1509.02971},
  archivePrefix = {arXiv}
}

@inproceedings{Fujimoto2018,
  author        = {Scott Fujimoto and Herke van Hoof and David Meger},
  title         = {Addressing function approximation error in actor-critic methods},
  booktitle     = {Proceedings of the 35th International Conference on Machine Learning},
  year          = {2018},
  volume        = {80},
  series        = {Proceedings of Machine Learning Research},
  pages         = {1587--1596}
}

@inproceedings{Haarnoja2018,
  author        = {Tuomas Haarnoja and Aurick Zhou and Pieter Abbeel and Sergey Levine},
  title         = {Soft actor-critic: off-policy maximum entropy deep reinforcement learning with a stochastic actor},
  booktitle     = {Proceedings of the 35th International Conference on Machine Learning},
  year          = {2018},
  volume        = {80},
  series        = {Proceedings of Machine Learning Research},
  pages         = {1861--1870}
}

@article{Brecht2015,
  author        = {Benjamin Brecht and Dileep V. Reddy and Christine Silberhorn and Michael G. Raymer},
  title         = {Photon temporal modes: a complete framework for quantum information science},
  journal       = {Physical Review X},
  year          = {2015},
  volume        = {5},
  number        = {4},
  pages         = {041017},
  doi           = {10.1103/PhysRevX.5.041017}
}

@article{Lukens2017,
  author        = {Joseph M. Lukens and Pavel Lougovski},
  title         = {Frequency-encoded photonic qubits for scalable quantum information processing},
  journal       = {Optica},
  year          = {2017},
  volume        = {4},
  number        = {1},
  pages         = {8--16},
  doi           = {10.1364/OPTICA.4.000008}
}

@article{Raymer2020,
  author        = {Michael G. Raymer and Ian A. Walmsley},
  title         = {Temporal modes in quantum optics: then and now},
  journal       = {Physica Scripta},
  year          = {2020},
  volume        = {95},
  number        = {6},
  pages         = {064002},
  doi           = {10.1088/1402-4896/ab6153}
}

@article{Jin2015,
  author        = {Rui-Bo Jin and Thomas Gerrits and Mikio Fujiwara and Ryota Wakabayashi and Taro Yamashita and Shigehito Miki and Hirotaka Terai and Ryosuke Shimizu and Masahide Sasaki},
  title         = {Spectrally resolved Hong--Ou--Mandel interference between independent photon sources},
  journal       = {Optics Express},
  year          = {2015},
  volume        = {23},
  number        = {22},
  pages         = {28836--28848},
  doi           = {10.1364/OE.23.028836}
}

@article{Yan2018,
  author        = {Yan, Fei and Krantz, Philip and Sung, Youngkyu and Kjaergaard, Morten and Campbell, Daniel L. and Wang, Joel I. J. and Orlando, Terry P. and Gustavsson, Simon and Oliver, William D.},
  title         = {{Tunable Coupling Scheme for Implementing High-Fidelity Two-Qubit Gates}},
  journal       = {Physical Review Applied},
  year          = {2018},
  volume        = {10},
  number        = {5},
  pages         = {054062},
  doi           = {10.1103/PhysRevApplied.10.054062},
  url           = {https://link.aps.org/doi/10.1103/PhysRevApplied.10.054062}
}

@article{Koch2007,
  author        = {Koch, Jens and Yu, Terri M. and Gambetta, Jay and Houck, A. A. and Schuster, D. I. and Majer, J. and Blais, Alexandre and Devoret, M. H. and Girvin, S. M. and Schoelkopf, R. J.},
  title         = {{Charge-insensitive qubit design derived from the Cooper pair box}},
  journal       = {Physical Review A},
  year          = {2007},
  volume        = {76},
  number        = {4},
  pages         = {042319},
  doi           = {10.1103/PhysRevA.76.042319},
  url           = {https://link.aps.org/doi/10.1103/PhysRevA.76.042319}
}

@article{Sung2021TunableCoupler,
  author        = {Sung, Youngkyu and Ding, Leon and Braum{\"u}ller, Jochen and Veps{\"a}l{\"a}inen, Antti and Kannan, Bharath and Kjaergaard, Morten and Greene, Ami and Samach, Gabriel O. and McNally, Chris and Kim, David and Melville, Alexander and Niedzielski, Bethany M. and Schwartz, Mollie E. and Yoder, Jonilyn L. and Orlando, Terry P. and Gustavsson, Simon and Oliver, William D.},
  title         = {{Realization of High-Fidelity CZ and ZZ-Free iSWAP Gates with a Tunable Coupler}},
  journal       = {Physical Review X},
  year          = {2021},
  volume        = {11},
  number        = {2},
  pages         = {021058},
  doi           = {10.1103/PhysRevX.11.021058},
  url           = {https://link.aps.org/doi/10.1103/PhysRevX.11.021058}
}

@article{Li2025PulseCalibration,
  author        = {Li, Tian-Ming and Zhang, Jia-Chi and Chen, Bing-Jie and Huang, Kaixuan and Liu, Hao-Tian and Xiao, Yong-Xi and Deng, Cheng-Lin and Liang, Gui-Han and Chen, Chi-Tong and Liu, Yu and Li, Hao and Bao, Zhen-Ting and Zhao, Kui and Xu, Yueshan and Li, Li and He, Yang and Liu, Zheng-He and Yu, Yi-Han and Zhou, Si-Yun and Liu, Yan-Jun and Song, Xiaohui and Zheng, Dongning and Xiang, Zhongcheng and Shi, Yun-Hao and Xu, Kai and Fan, Heng},
  title         = {{High-precision pulse calibration of tunable couplers for high-fidelity two-qubit gates in superconducting quantum processors}},
  journal       = {Physical Review Applied},
  year          = {2025},
  volume        = {23},
  number        = {2},
  pages         = {024059},
  doi           = {10.1103/PhysRevApplied.23.024059},
  url           = {https://link.aps.org/doi/10.1103/PhysRevApplied.23.024059}
}

@article{AutoCalib112Qubit,
  author        = {Xu, Huikai and Han, Jiaxiu and Ou, Shigang and Ye, Cheng and Shen, Zisong and Gao, Jing and Wang, Yijia and Che, Tianrui and Song, Yu and Liu, Weiyang and Wang, Lei and Zhang, Lin-Feng and Zhang, Pan and Yu, Hai-Feng},
  title         = {{Vibe Calibration: Autonomous Bring-up of a 112-Qubit Superconducting Quantum Processor by a Skill-Orchestrating Language Agent}},
  journal       = {arXiv: 2606.22376},
  year          = {2026},
  url           = {https://arxiv.org/abs/2606.22376},
  eprint        = {2606.22376},
  archivePrefix = {arXiv}
}

@article{ScalingCalibrationReport,
  author        = {Mohseni, Masoud and Scherer, Artur and Johnson, K. Grace and Wertheim, Oded and Otten, Matthew and Anand, Namit and Aadit, Navid Anjum and Alexeev, Yuri and Ben-Shach, Gilad and Bresniker, Kirk M. and Camsari, Kerem Y. and Chapman, Barbara and Chatterjee, Soumitra and Chowdhury, Shuvro and others},
  title         = {{How to Build a Quantum Supercomputer: Scaling from Hundreds to Millions of Qubits}},
  journal       = {arXiv: 2411.10406},
  year          = {2024},
  url           = {https://arxiv.org/abs/2411.10406},
  eprint        = {2411.10406},
  archivePrefix = {arXiv}
}

@book{Mandel1995,
  author        = {Mandel, Leonard and Wolf, Emil},
  title         = {Optical Coherence and Quantum Optics},
  publisher     = {Cambridge University Press},
  year          = {1995},
  address       = {Cambridge}
}

@book{Ou2007,
  author        = {Ou, Zhe-Yu Jeff},
  title         = {Multi-Photon Quantum Interference},
  publisher     = {Springer},
  year          = {2007},
  doi           = {10.1007/978-0-387-25554-5},
  url           = {https://doi.org/10.1007/978-0-387-25554-5}
}

@article{Gerrits2015PRA,
  author        = {Gerrits, Thomas and Marsili, Francesco and Verma, Varun B. and Shalm, Lynden K. and Shaw, Matthew D. and Mirin, Richard P. and Nam, Sae Woo},
  title         = {Spectral correlation measurements at the {H}ong-{O}u-{M}andel interference dip},
  journal       = {Physical Review A},
  year          = {2015},
  volume        = {91},
  pages         = {013830},
  doi           = {10.1103/PhysRevA.91.013830}
}

@article{YepizGraciano2020PR,
  author        = {Yepiz-Graciano, Pablo and Angulo Mart{\'\i}nez, Al{\'\i} Michel and Lopez-Mago, Dorilian and Cruz-Ramirez, Hector and U'Ren, Alfred B.},
  title         = {Spectrally resolved {H}ong--{O}u--{M}andel interferometry for quantum-optical coherence tomography},
  journal       = {Photonics Research},
  year          = {2020},
  volume        = {8},
  number        = {6},
  pages         = {1023--1034},
  doi           = {10.1364/PRJ.388693}
}

@article{Lavoie2013,
  author        = {Lavoie, J. and Donohue, J. M. and Wright, L. G. and Fedrizzi, A. and Resch, K. J.},
  title         = {{Spectral compression of single photons}},
  journal       = {Nature Photonics},
  year          = {2013},
  volume        = {7},
  number        = {5},
  pages         = {363--366},
  doi           = {10.1038/nphoton.2013.47},
  url           = {http://dx.doi.org/10.1038/nphoton.2013.47 http://www.nature.com/doifinder/10.1038/nphoton.2013.47}
}

@article{Karpinski2017,
  author        = {Karpi{\'{n}}ski, Micha{\l} and Jachura, Micha{\l} and Wright, Laura J. and Smith, Brian J.},
  title         = {{Bandwidth manipulation of quantum light by an electro-optic time lens}},
  journal       = {Nature Photonics},
  year          = {2017},
  volume        = {11},
  number        = {1},
  pages         = {53--57},
  doi           = {10.1038/nphoton.2016.228},
  url           = {https://www.nature.com/articles/nphoton.2016.228}
}

@article{Sosnicki2023,
  author        = {So{\'{s}}nicki, Filip and Miko{\l}ajczyk, Micha{\l} and Golestani, Ali and Karpi{\'{n}}ski, Micha{\l}},
  title         = {{Interface between picosecond and nanosecond quantum light pulses}},
  journal       = {Nature Photonics},
  year          = {2023},
  volume        = {17},
  number        = {9},
  pages         = {761--766},
  doi           = {10.1038/s41566-023-01214-z},
  url           = {https://www.nature.com/articles/s41566-023-01214-z}
}

@article{Lu23,
  author        = {Hsuan-Hao Lu and Marco Liscidini and Alexander L. Gaeta and Andrew M. Weiner and Joseph M. Lukens},
  title         = {Frequency-bin photonic quantum information},
  journal       = {Optica},
  year          = {2023},
  volume        = {10},
  number        = {12},
  pages         = {1655--1671},
  doi           = {10.1364/OPTICA.506096},
  url           = {https://opg.optica.org/optica/abstract.cfm?URI=optica-10-12-1655}
}

@article{Buddhiraju2021,
  author        = {Buddhiraju, Siddharth and Dutt, Avik and Minkov, Momchil and Williamson, Ian A. D. and Fan, Shanhui},
  title         = {{Arbitrary linear transformations for photons in the frequency synthetic dimension}},
  journal       = {Nature Communications},
  year          = {2021},
  volume        = {12},
  number        = {1},
  pages         = {2401},
  doi           = {10.1038/s41467-021-22670-7},
  url           = {https://www.nature.com/articles/s41467-021-22670-7}
}

@article{Zhu2022,
  author        = {Zhu, Di and Chen, Changchen and Yu, Mengjie and Shao, Linbo and Hu, Yaowen and Xin, C. J. and Yeh, Matthew and Ghosh, Soumya and He, Lingyan and Reimer, Christian and Sinclair, Neil and Wong, Franco N. C. and Zhang, Mian and Lon{\v{c}}ar, Marko},
  title         = {{Spectral control of nonclassical light pulses using an integrated thin-film lithium niobate modulator}},
  journal       = {Light: Science \& Applications},
  year          = {2022},
  volume        = {11},
  number        = {1},
  pages         = {327},
  doi           = {10.1038/s41377-022-01029-7},
  url           = {https://www.nature.com/articles/s41377-022-01029-7}
}

@article{Yang2026,
  author        = {Yang, Ran and Zhou, Wei and Guo, Dong-Jie and Ke, Hong-Ming and Tao, Linrunde and Wei, Ying and Duan, Jia-Chen and Cui, Yu and Jia, Kunpeng and Xie, Zhenda and Lin, Zhongjin and Cai, Xinlun and Gong, Yan-Xiao and Zhu, Shi-Ning},
  title         = {{Quantum photonic frequency processor on thin-film lithium niobate}},
  journal       = {arXiv: 2603.11471},
  year          = {2026},
  url           = {http://arxiv.org/abs/2603.11471},
  eprint        = {2603.11471},
  archivePrefix = {arXiv}
}

@article{Hoeffding1963,
  author        = {Hoeffding, Wassily},
  title         = {Probability Inequalities for Sums of Bounded Random Variables},
  journal       = {Journal of the American Statistical Association},
  year          = {1963},
  volume        = {58},
  number        = {301},
  pages         = {13--30},
  doi           = {10.1080/01621459.1963.10500830}
}

@misc{brockman2016openai,
  author        = {Greg Brockman and Vicki Cheung and Ludwig Pettersson and Jonas Schneider and John Schulman and Jie Tang and Wojciech Zaremba},
  title         = {OpenAI Gym},
  year          = {2016},
  doi           = {10.48550/ARXIV.1606.01540},
  url           = {https://arxiv.org/abs/1606.01540},
  eprint        = {1606.01540},
  archivePrefix = {arXiv}
}

@article{McKay2017EfficientZ,
  author        = {McKay, David C. and Wood, Christopher J. and Sheldon, Sarah
            and Chow, Jerry M. and Gambetta, Jay M.},
  title         = {Efficient {Z} gates for quantum computing},
  journal       = {Phys. Rev. A},
  year          = {2017},
  volume        = {96},
  pages         = {022330},
  doi           = {10.1103/PhysRevA.96.022330}
}

@article{Schuld2019,
  author        = {Maria Schuld and Nathan Killoran},
  title         = {Quantum Machine Learning in Feature Hilbert Spaces},
  journal       = {Physical Review Letters},
  year          = {2019},
  volume        = {122},
  pages         = {040504},
  url           = {https://journals.aps.org/prl/abstract/10.1103/PhysRevLett.122.040504}
}
\clearpage

\clearpage
\onecolumngrid

\setcounter{section}{0}
\setcounter{equation}{0}
\setcounter{figure}{0}
\setcounter{table}{0}
\setcounter{theorem}{0}
\setcounter{remark}{0}
\renewcommand{\theequation}{S\arabic{equation}}
\renewcommand{\thefigure}{S\arabic{figure}}
\renewcommand{\thetable}{S\arabic{table}}
\renewcommand{\thetheorem}{S\arabic{theorem}}

\renewcommand{\theHequation}{supp.equation.\arabic{equation}}
\renewcommand{\theHfigure}{supp.figure.\arabic{figure}}
\renewcommand{\theHtable}{supp.table.\arabic{table}}

\begin{center}
  {\large\bfseries Supplemental Material for}\\[0.5em]
  {\large\itshape Quantum Optical Reinforcement Learning via Spectrum-Resolved Hong-Ou-Mandel Interference}
\end{center}
\vspace{1em}

\section{Numerical implementation of the SR-HOM agent}

This section describes how the spectrum-resolved Hong-Ou-Mandel (SR-HOM) readout is implemented numerically using a finite-bin spectral representation, and how the resulting spectral features are used to construct the actor and critic networks in the reinforcement-learning framework.

\subsection{Finite-bin spectral representation}

For numerical simulation, each continuous spectral mode is discretized into
$K$ frequency bins. A real-valued environment observation
$s\in\mathbb{R}^{d}$ is encoded into a normalized complex spectral-amplitude vector,
\begin{equation}
\boldsymbol{\psi}_{\theta}(s)
=
\bigl(
\psi_{\theta,1}(s),\ldots,\psi_{\theta,K}(s)
\bigr)
\in\mathbb{C}^{K},
\qquad
\sum_{i=1}^{K}
\left|\psi_{\theta,i}(s)\right|^{2}
=1,
\end{equation}
where $\theta$ denotes the trainable parameters of the spectral encoder. The normalization ensures that $\boldsymbol{\psi}_{\theta}(s)$ represents a valid single-photon spectral state in the discretized frequency basis.

Analogously, the trainable probe spectra are parameterized in the same finite-bin basis
and normalized before the spectrum-resolved Hong-Ou-Mandel readout is evaluated. This discretization provides a finite-dimensional approximation to the continuous spectral wavefunctions introduced in the main text and converges to the continuous description as the frequency-bin resolution is increased.

\subsection{Spectrum-resolved coincidence probability matrix $C_{ij}$}

In the main text, \(C_{ij}\) denotes the conditional probability that the two detected photons occupy frequency bins \(\Omega_i\) and \(\Omega_j\), respectively, given that a coincidence event has occurred. The sample space is therefore the set of accepted coincidence events. Let \(P_{\cd}(\omega_1,\omega_2;\bm s,\lambda)\) be the corresponding normalized frequency-resolved coincidence density, where \(\bm s\) is the encoded environment observation and \(\bm \lambda\) collectively denotes the trainable optical parameters. The finite-bin coincidence probability matrix is then 
\begin{equation} 
C_{ij}(\bm s,\bm \lambda) = \int_{\Omega_i} d\omega_1 \int_{\Omega_j} d\omega_2\, P_{cd}(\omega_1,\omega_2;\bm s,\bm\lambda), \qquad \sum_{i,j}C_{ij}(\bm s,\bm\lambda)=1. \label{eq:supp_coincidence_matrix} 
\end{equation}

In a photon-counting experiment, suppose that \(n_{ij}\) coincidence events are registered in the bin pair \((i,j)\), and let \(M=\sum_{i,j}n_{ij}\) be the total number of accepted coincidence events. Then the empirical estimator of Eq.~\eqref{eq:supp_coincidence_matrix} can be written as: 
\begin{equation} 
\widehat{C}_{ij} = \frac{n_{ij}}{M}. 
\end{equation} 
The overall HOM coincidence probability does not affect the normalization of \(C_{ij}\). Instead, it determines the number of photon-pair trials required to collect a prescribed number \(M\) of accepted coincidence samples and therefore enters the photon-pair cost appearing in the sampling bound of the main text. In the numerical simulations, \(C_{ij}\) is evaluated directly from the normalized spectral-mode coefficients in the exact-probability limit. Finite-shot photon-counting fluctuations are therefore not included in the reported learning curves. In a hardware implementation, the exact entries would be replaced by the empirical estimates \(\widehat{C}_{ij}\), with the associated sampling cost and statistical error characterized by the bound given in the main text.

\subsection{Actor readout}

For an \(m\)-dimensional continuous-action space, the frequency bins are
partitioned into \(m\) disjoint groups,
\(\{\mathcal{G}_r\}_{r=1}^{m}\). The actor used in the numerical benchmarks
assigns one grouped SR-HOM response to each action component, following the
diagonal-readout construction introduced in the main text. We first define
the same-bin coincidence probabilities
\begin{equation}
p_i(\bm s,\bm \lambda)
=
C_{ii}(\bm s,\bm \lambda).
\end{equation}
The optical response associated with the \(r\)th action component is then
\begin{equation}
z_r(\bm s,\bm \lambda)
=
\sum_{i\in\mathcal{G}_r}p_i(\bm s,\bm \lambda),
\qquad
\mu_r(s)
=
g_r\!\left[z_r(\bm s,\bm \lambda)\right],
\label{eq:supp_actor_readout}
\end{equation}
where \(g_r\) is a fixed monotone transformation that maps the nonnegative SR-HOM response to the admissible interval of the \(r\)th environmental control variable. This final transformation determines both the scale and sign of the physical control signal.

During training, \(\mu_r(s)\) serves as the mean of the exploration policy in RL.
The policy standard deviation is parameterized by a trainable
log-standard-deviation vector and controls the amount of stochastic
exploration. Sampled actions are subsequently constrained to the allowed
action range of the environment. Thus, the trainable optical component of the
actor consists of one grouped spectral response for each action dimension,
whereas the Gaussian exploration layer is a standard device used only for
continuous-control policy optimization.

At evaluation time, exploration noise is removed and the deterministic mean
action $a_r(s)=\mu_r(s)$
is applied to the environment.

\subsection{Critic readout}

The critic employs the same finite-bin SR-HOM principle to map a state-action
pair to a scalar value estimate. The state and action are first encoded into
normalized complex spectral vectors,
\begin{equation}
\boldsymbol{\alpha}_{\eta_{\bm{s}}}(\bm s)\in\mathbb{C}^{K},
\qquad
\boldsymbol{\beta}_{\eta_a}(\bm a)\in\mathbb{C}^{L},
\end{equation}
satisfying
\begin{equation}
\sum_{i=1}^{K}
\left|\alpha_{\eta_s,i}(\bm s)\right|^2=1,
\qquad
\sum_{j=1}^{L}
\left|\beta_{\eta_a,j}(\bm a)\right|^2=1.
\end{equation}
Here, \(\eta_s\) and \(\eta_a\) denote the trainable parameters of the state and action encoders, respectively.

The two encoded spectra are combined through normalized trainable probe
spectra
\(\boldsymbol{u}\in\mathbb{C}^{K}\) and
\(\boldsymbol{v}\in\mathbb{C}^{L}\). A compact separable implementation of the
critic features takes the form
\begin{equation}
F_i^{Q}(\bm s,\bm a)
\propto
\left|\alpha_{\eta_s,i}(s)\right|^2
\left|\bm u_i\right|^2
\left|
\left\langle
\boldsymbol{v},
\boldsymbol{\beta}_{\eta_a}(a)
\right\rangle
\right|^2,
\label{eq:supp_critic_features}
\end{equation}
with normalized probe spectra $\bm u$ and $\bm v$.
The first two factors describe the bin-resolved response of the encoded state
to the state probe, whereas the final overlap introduces an
action-dependent modulation. 

The bin-resolved features are subsequently collected into a small number of
disjoint groups \(\{\mathcal{B}_{\ell}\}\) and mapped linearly to a scalar
critic output,
\begin{equation}
Q(\bm s,\bm a)
=
b_Q
+
\sum_{\ell}
W_{\ell}
\sum_{i\in\mathcal{B}_{\ell}}
F_i^{Q}(\bm s,\bm a),
\label{eq:supp_critic_readout}
\end{equation}
where \(W_{\ell}\) and \(b_Q\) are trainable classical readout parameters.
We can also define the spectrum-resolved readout tensor similar to the actor
\begin{equation}
q_i(\bm{x}, \bm{\kappa})
=
C_{ii}(\bm{x}, \bm{\kappa}),
\end{equation}
where $\bm{\kappa}$ denotes the trainable parameters of the SR-HOM module and input $\bm x$ here refers to action-state pair $(\bm s, \bm a)$. The resulting tensor is mapped to a scalar action-value estimate through a trainable linear readout,
\begin{equation}
Q(\bm{x}, \bm{\kappa})
=
f_{\bm{W},b}
\left(
\sum_{i} q_i(\bm{x}, \bm{\kappa})
\right),
\label{eq:supp_q_abstract}
\end{equation}
where $f_{\bm{W},b}$ is parameterized by trainable weights $\bm{W}$ and
bias $b$.
The separable construction keeps the value estimator
compact while allowing the action encoding to modulate the
state-dependent spectral response.

\subsection{Training protocol}

The SR-HOM and multilayer perceptron (MLP) agents are trained using the same PPO-style on-policy
actor--critic protocol. At each training iteration, the current policy is
used to collect a batch of environment trajectories. The observed rewards
and critic predictions are then used to construct discounted-return or
advantage estimates. The actor parameters are updated using a clipped
policy-gradient objective, which limits excessively large changes in the
policy between successive iterations, while the critic is trained by
minimizing a squared-error loss relative to the corresponding return target.

The optimization protocol itself is independent of the optical
implementation. In the architectural comparison, the conventional MLP actor
and critic are replaced by the SR-HOM function approximators described above,
while the environment interaction procedure and the high-level
reinforcement-learning protocol are kept fixed. Consequently, the comparison
isolates the effect of the SR-HOM architecture rather than changes in the
underlying policy-optimization algorithm.

\subsection{Benchmark protocol}

We evaluate the proposed spectrum-resolved Hong-Ou-Mandel (SR-HOM) reinforcement-learning agent against parameter-matched MLP baselines across five continuous-control environments. The MLP baseline uses the same observation and action spaces and is trained with the same bounded Gaussian actor--critic procedure as the SR-HOM agent. The only architectural difference is that the SR-HOM actor and critic modules are replaced by compact two-hidden-layer neural networks constructed under the same parameter-budget constraint adopted in the main-text comparison. The MLP actor outputs the mean of the continuous action distribution, whereas the critic takes the concatenated state--action pair as input and returns a scalar value estimate.

For each environment, candidate MLP configurations were selected through a hyperparameter search over network widths, learning rates, and exploration scales. Here, an episode denotes one complete interaction trajectory, beginning from an environment reset and ending when the task terminates, or the prescribed maximum length is reached. The selected configurations were then evaluated across multiple random seeds, and the baseline reported for each task was chosen according to its highest 100-episode moving-average return. This procedure deliberately favors the MLP baseline, because failed or unstable configurations are excluded from the primary comparison. To ensure a consistent evaluation window, all reported metrics are computed after truncating the SR-HOM and MLP return sequences to their common episode length. The environment specifications are summarized in Table~\ref{tab:supp_env_specs}.

\section{Additional quantitative benchmark results}

Table~\ref{tab:supp_metrics_all_envs} reports the quantitative comparison between SR-HOM and the parameter-matched MLP baseline. We report four metrics: the first episode at which the raw episodic return reaches the environment-specific threshold, the post-threshold collapse rate, the curve volatility measured relative to a local moving-average trend, and the best 100-episode moving-average return. These metrics quantify sample efficiency, post-threshold stability, short-scale return fluctuations, and peak smoothed performance, respectively.

\begin{table*}[t]
\centering
\caption{
Environment specifications for the five continuous-control benchmarks.
The environments are sorted by observation dimensionality.
}
\label{tab:supp_env_specs}

{\scriptsize
\setlength{\tabcolsep}{2.2pt}
\renewcommand{\arraystretch}{0.95}

\begin{tabular*}{\textwidth}{@{}p{2.15cm}@{\extracolsep{\fill}}cccp{8.7cm}@{}}
\toprule
Environment 
& \makecell[c]{Observation\\dim.}
& \makecell[c]{Action\\dim.}
& Max episode length
& Reward threshold and environment specification note \\
\midrule

\makecell[l]{LunarLander\\Continuous-v3}
& 8
& 2
& 1000
& Solved threshold $=200$. The reward includes landing bonuses, leg-contact rewards, crash penalties, and engine-use penalties, so there is no simple closed-form maximum. \\

\makecell[l]{InvertedDouble\\Pendulum-v5}
& 9
& 1
& 1000
& Threshold $=9100$. The approximate maximum is $\sim 10000$, corresponding to near-maximal per-step survival reward over 1000 steps with minimal penalties. \\

\makecell[l]{Reacher-v5}
& 10
& 2
& 50
& Threshold $=-3.75$. The reward is non-positive and consists mainly of distance and control penalties; the approximate maximum is $0$. \\

\makecell[l]{Bipedal\\Walker-v3}
& 24
& 4
& 1600
& Solved threshold $=300$. The return depends on forward progress, terrain completion, falling penalty, and torque cost; a strict closed-form maximum is not typically reported. \\

\makecell[l]{Ant-v5}
& 105
& 8
& 1000
& Reference threshold $=5000$. The reward includes forward velocity, healthy reward, control cost, and contact cost; therefore no simple finite closed-form maximum is available. \\

\bottomrule
\end{tabular*}
}

\end{table*}

\begin{table*}[t]
\centering
\caption{
Quantitative comparison between the proposed SR-HOM agent and the
parameter-matched MLP baseline across five continuous-control benchmarks.
For each environment, the return threshold or reference return level is
shown in parentheses. ``Episodes to threshold'' denotes the first episode
at which the raw episodic return reaches the environment-specific threshold.
The collapse rate is the fraction of subsequent episodes for which the raw
return falls below one-half of the corresponding threshold, or double of the threshold for negative thresholds. It therefore
quantifies severe post-threshold performance degradation. Curve volatility
is defined as the standard deviation of the residual fluctuations obtained
after subtracting the local moving-average trend from the raw return
trajectory. The best 100-episode moving-average return is the maximum value
of the 100-episode simple moving average attained over the aligned training
horizon. For Ant-v5, the value 5000 is used as a reference return level
rather than a universal solved criterion. N.R. indicates that the threshold
was \emph{not reached} within the aligned training horizon, whereas N.D. indicates
that the collapse rate is \emph{not defined} because the corresponding post-threshold
regime was not reached. The arrows indicate whether higher ($\uparrow$) or lower ($\downarrow$) values are preferable.
}
\label{tab:supp_metrics_all_envs}

{\scriptsize
\setlength{\tabcolsep}{2.2pt}
\renewcommand{\arraystretch}{0.92}

\begin{tabular*}{\textwidth}{@{}p{2.35cm}@{\extracolsep{\fill}}rrrrrrrr@{}}
\toprule
Environment
& \multicolumn{2}{c}{\makecell[c]{Episodes to\\threshold ($\downarrow$)}}
& \multicolumn{2}{c}{\makecell[c]{Collapse\\rate ($\downarrow$)}}
& \multicolumn{2}{c}{\makecell[c]{Curve\\volatility}}
& \multicolumn{2}{c@{}}{\makecell[c]{Best 100-ep\\moving-avg. return ($\uparrow$)}} \\
\cmidrule(lr){2-3}
\cmidrule(lr){4-5}
\cmidrule(lr){6-7}
\cmidrule(l){8-9}
& \multicolumn{1}{c}{SR-HOM}
& \multicolumn{1}{c}{MLP}
& \multicolumn{1}{c}{SR-HOM}
& \multicolumn{1}{c}{MLP}
& \multicolumn{1}{c}{SR-HOM}
& \multicolumn{1}{c}{MLP}
& \multicolumn{1}{c}{SR-HOM}
& \multicolumn{1}{c@{}}{MLP} \\
\midrule

\makecell[l]{\textit{LunarLander}\\\textit{Continuous-v3}\\{\footnotesize $(200)$}}
& 666 & 2935
& 0.0881 & 0.4982
& 101.17 & 116.51
& 276.81 & 159.06 \\

\makecell[l]{\textit{InvertedDouble}\\\textit{Pendulum-v5}\\{\footnotesize $(9100)$}}
& 35203 & N.R.
& 0.0585 & N.D.
& 1377.53 & 92.63
& 9352.23 & 1000.00 \\

\makecell[l]{\textit{Reacher-v5}\\{\footnotesize $(-3.75)$}\\}
& 20362 & 48195
& 0.5597 & 0.9128
& 1.87 & 2.39
& $-3.44$ & $-4.47$ \\

\makecell[l]{\textit{Bipedal}\\\textit{Walker-v3}\\{\footnotesize $(300)$}}
& 8796 & N.R.
& 0.0349 & N.D.
& 49.62 & 34.37
& 313.09 & 11.55 \\

\makecell[l]{\textit{Ant-v5}\\{\footnotesize reference}\\{\footnotesize $(5000)$}}
& 17594 & N.R.
& 0.0585 & N.D.
& 883.10 & 406.38
& 5240.10 & 358.03 \\

\bottomrule
\end{tabular*}
}

\end{table*}

\section{Additional benchmark training curves}

\begin{figure*}[t]
   \centering

   \begin{minipage}{0.49\textwidth}
       \centering
       \includegraphics[width=\linewidth]{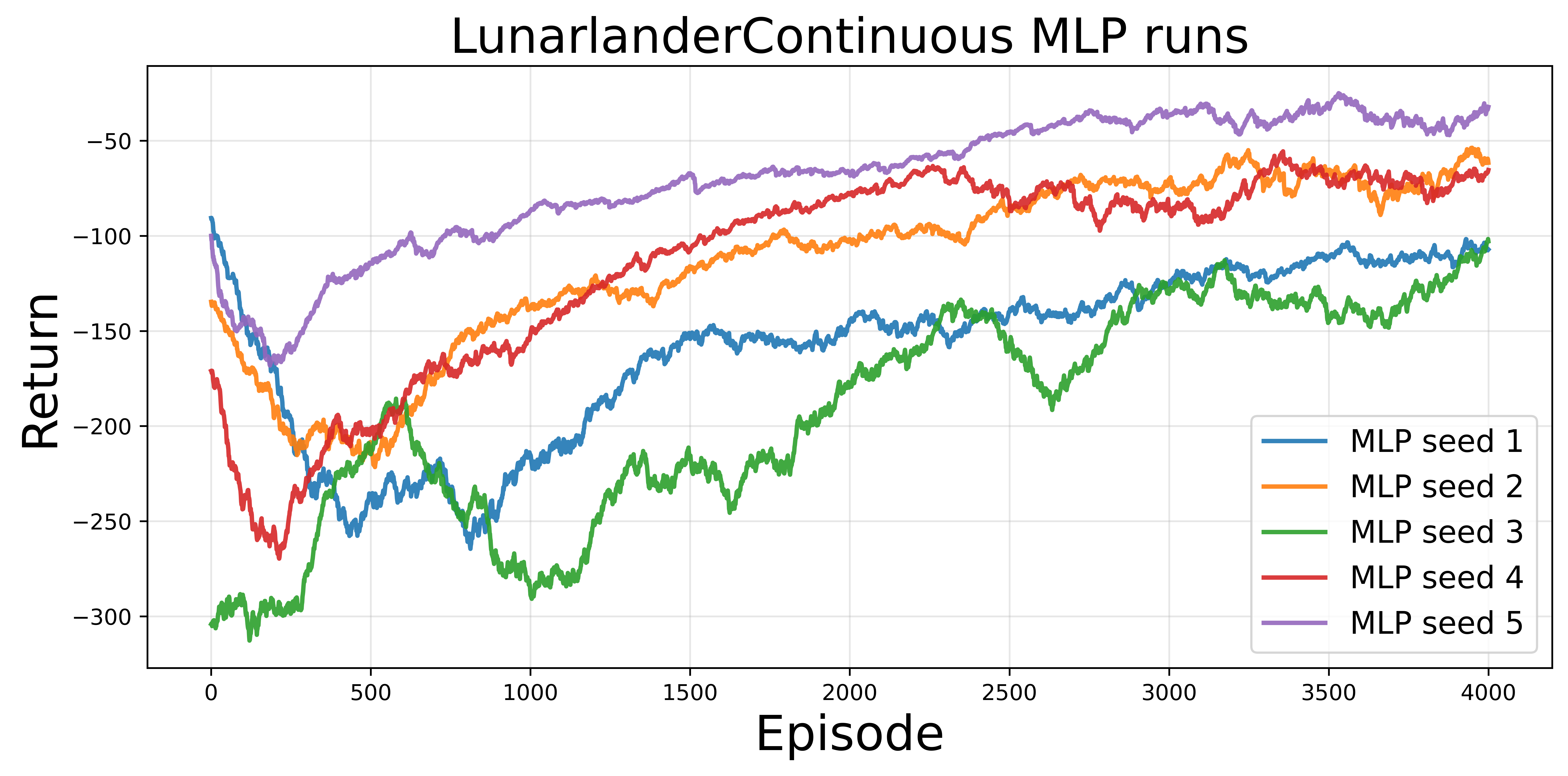}
       \vspace{-2mm}
       \panelcaption{(a) LunarLanderContinuous-v3}
   \end{minipage}
   \hfill
   \begin{minipage}{0.49\textwidth}
       \centering
       \includegraphics[width=\linewidth]{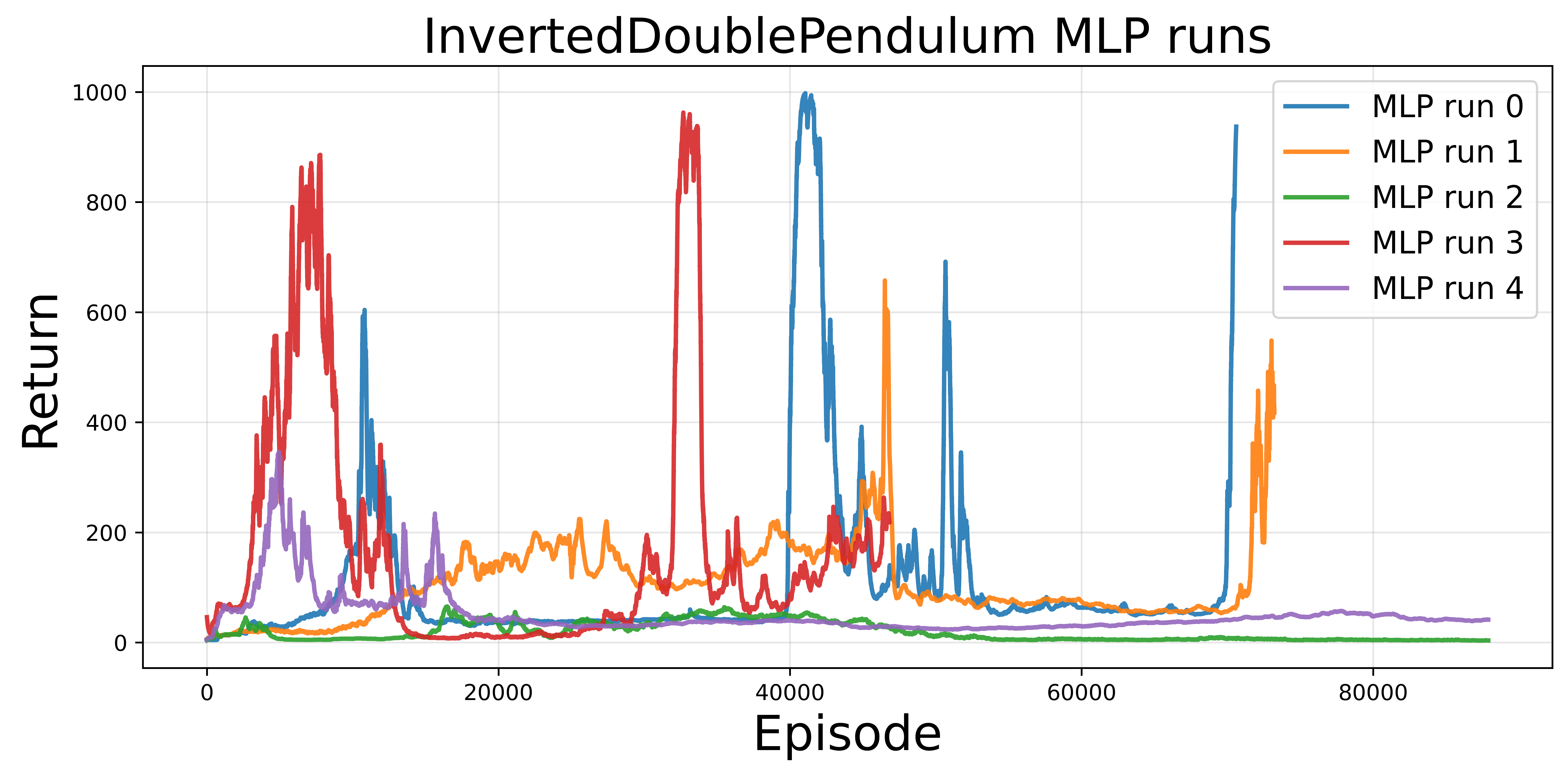}
       \vspace{-2mm}
       \panelcaption{(b) InvertedDoublePendulum-v5}
   \end{minipage}

   \vspace{2mm}

   \begin{minipage}{0.49\textwidth}
       \centering
       \includegraphics[width=\linewidth]{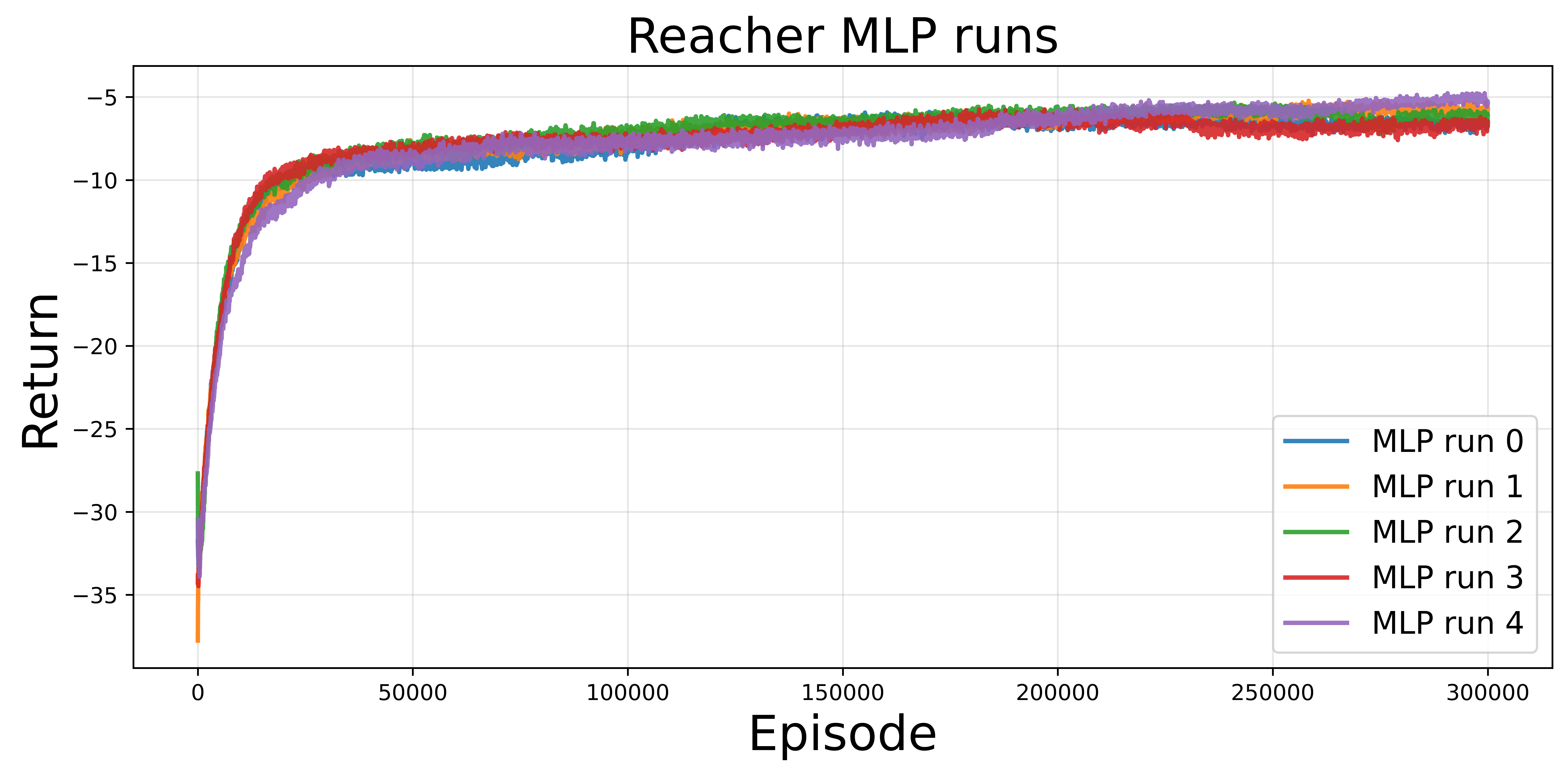}
       \vspace{-2mm}
       \panelcaption{(c) Reacher-v5}
   \end{minipage}
   \hfill
   \begin{minipage}{0.49\textwidth}
       \centering
       \includegraphics[width=\linewidth]{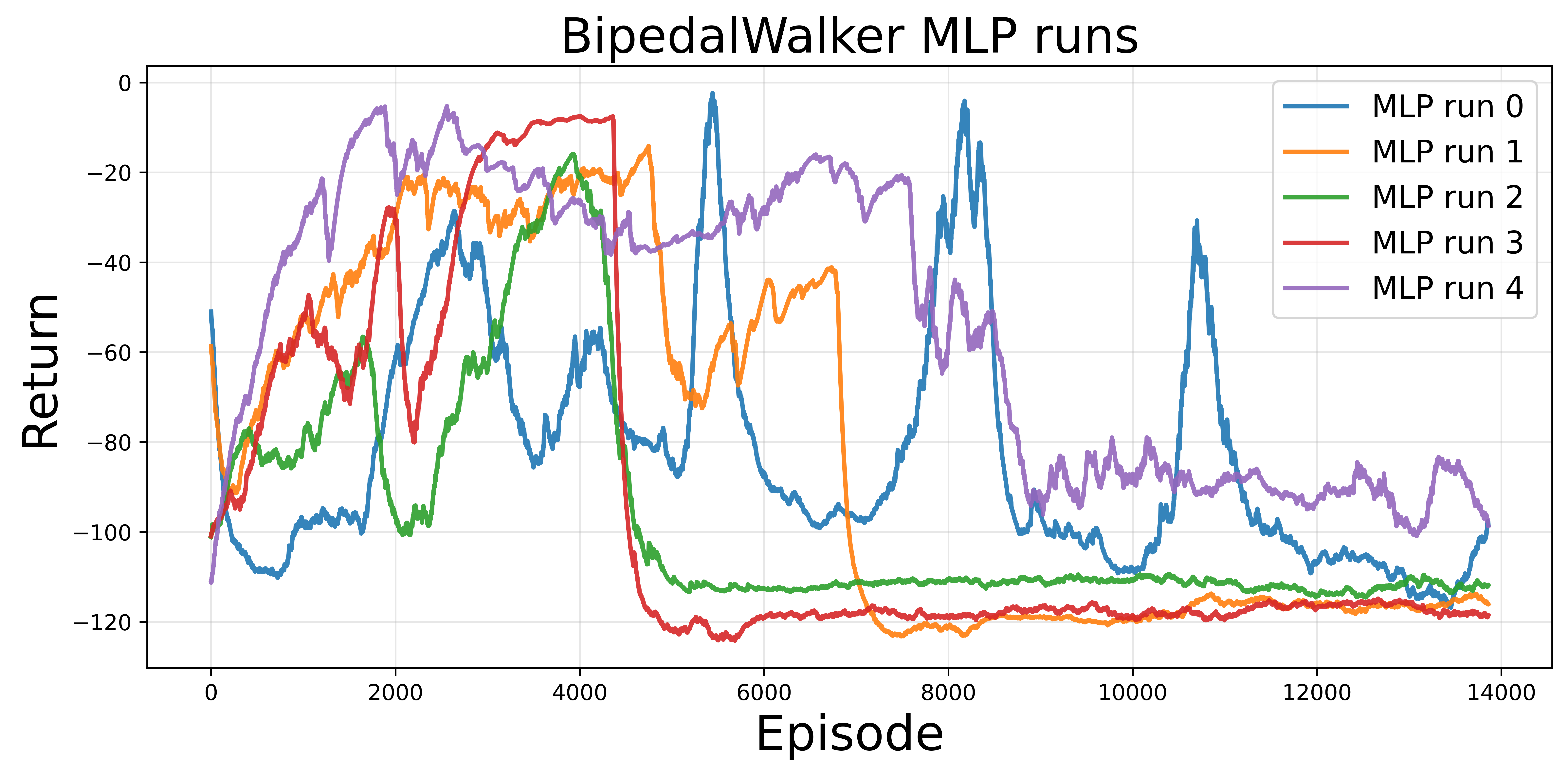}
       \vspace{-2mm}
       \panelcaption{(d) BipedalWalker-v3}
   \end{minipage}

   \vspace{2mm}

   \begin{minipage}{0.62\textwidth}
       \centering
       \includegraphics[width=\linewidth]{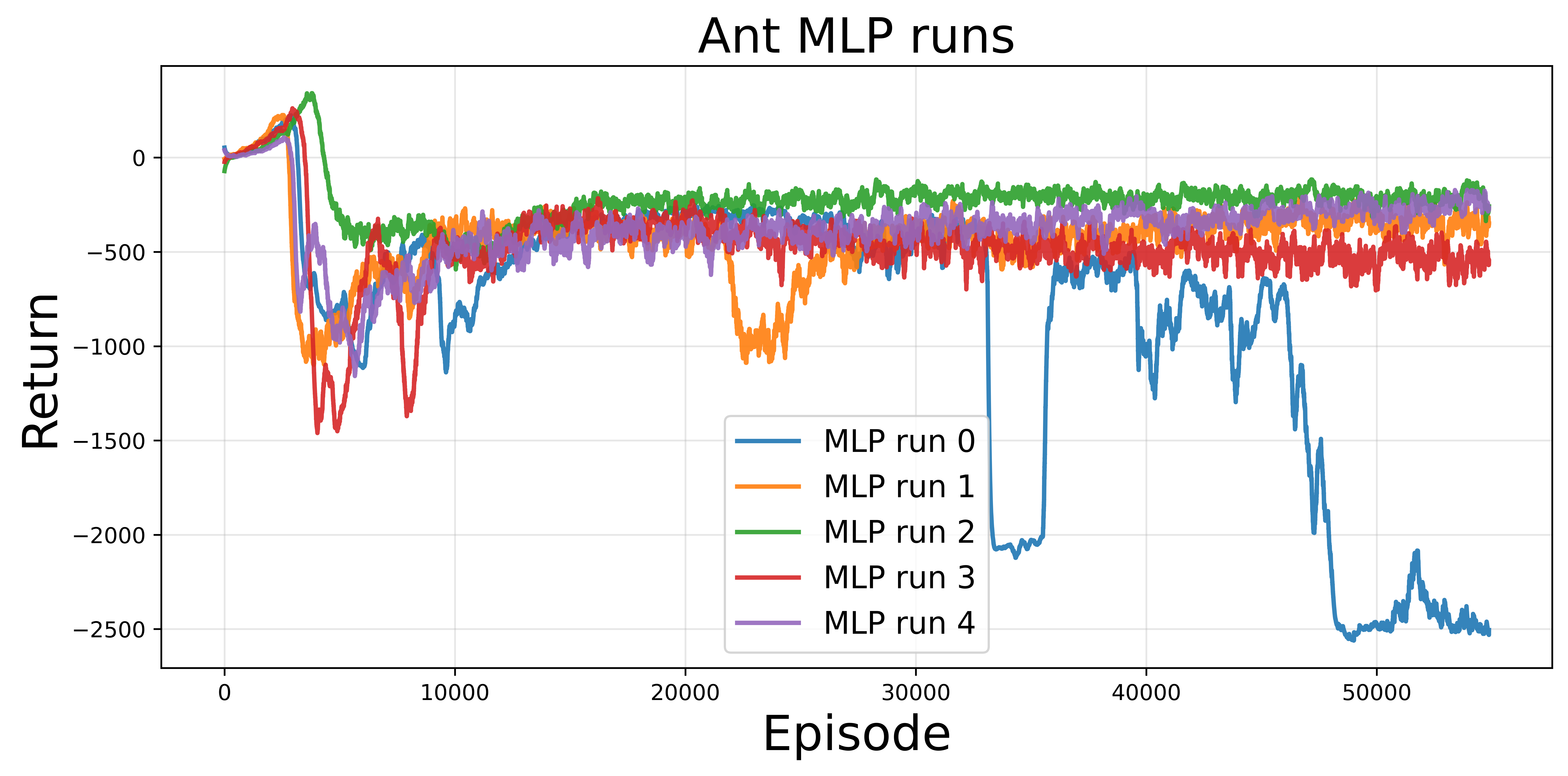}
       \vspace{-2mm}
       \panelcaption{(e) Ant-v5}
   \end{minipage}

   \caption{
   Multi-seed smoothed training-return curves for the parameter-matched MLP baselines.
   Each panel shows the exponentially smoothed episodic return trajectories from multiple random seeds.
   These curves illustrate the variability of the MLP baseline across independent training runs.
   For each benchmark, the MLP run used in the SR-HOM comparison is selected after hyperparameter search and multiple-seed evaluation according to the best 100-episode moving-average return, making the reported baseline favorable to the MLP rather than to failed or unstable runs.
   }
   \label{fig:supp_mlp_multiseed_curves}
\end{figure*}

Fig.~\ref{fig:supp_mlp_multiseed_curves} presents the training-return trajectories obtained by training the hyperparameter-selected, parameter-matched MLP configuration with multiple random seeds for each of the five continuous-control benchmarks.
These curves quantify the seed-to-seed variability of the neural baseline and make the baseline-selection procedure explicit. For each environment, a hyperparameter search is first performed to identify the best-performing parameter-matched MLP configuration. This configuration is then fixed and independently trained across multiple random seeds, and the resulting run with the highest 100-episode moving-average return is used for the comparison in the main text.
The full set of multi-seed trajectories therefore provides the context needed to assess the stability of the MLP baseline and to interpret its comparison with the SR-HOM agent.

\begin{figure}[!htbp]
   \centering

   \begin{minipage}{0.49\textwidth}
       \centering
       \includegraphics[width=\linewidth]{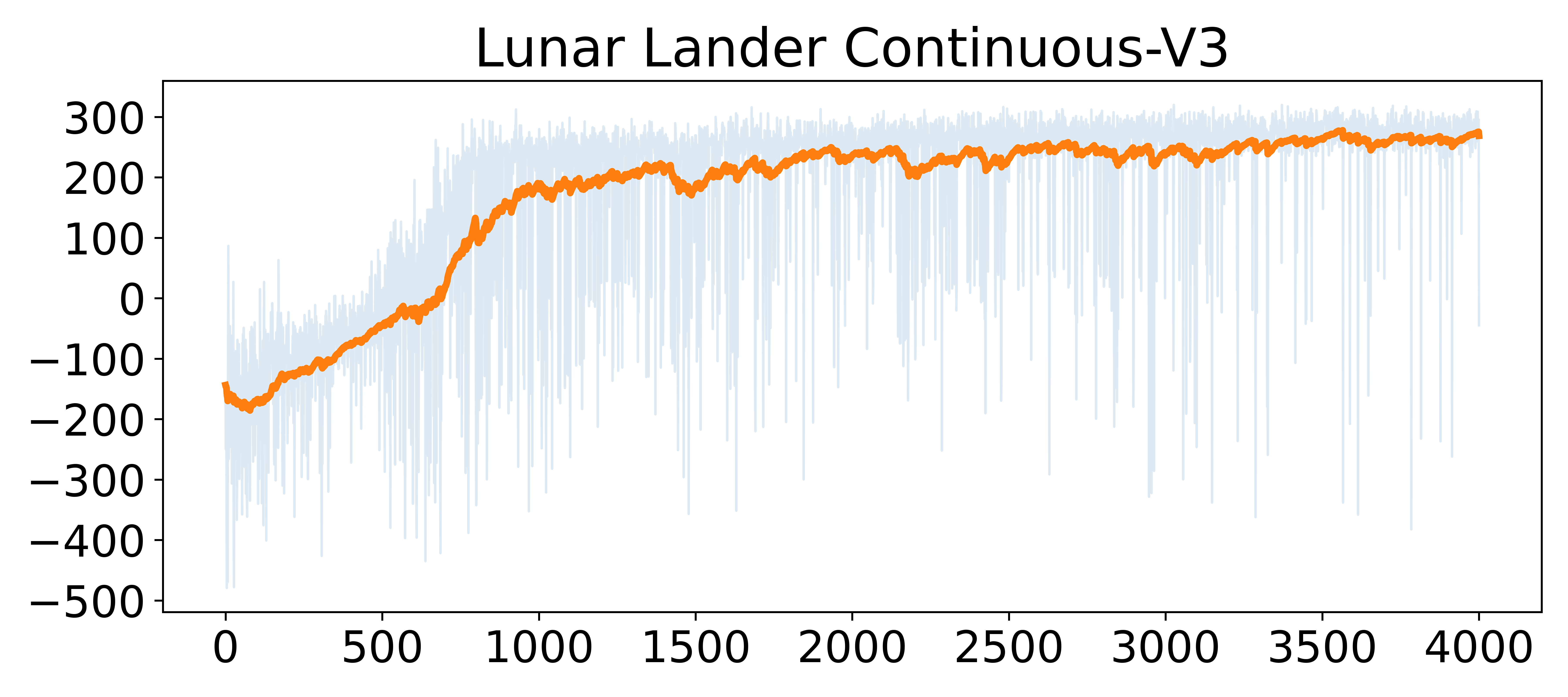}
       \vspace{-2mm}
       \panelcaption{(a) LunarLanderContinuous-v3}
   \end{minipage}
   \hfill
   \begin{minipage}{0.49\textwidth}
       \centering
       \includegraphics[width=\linewidth]{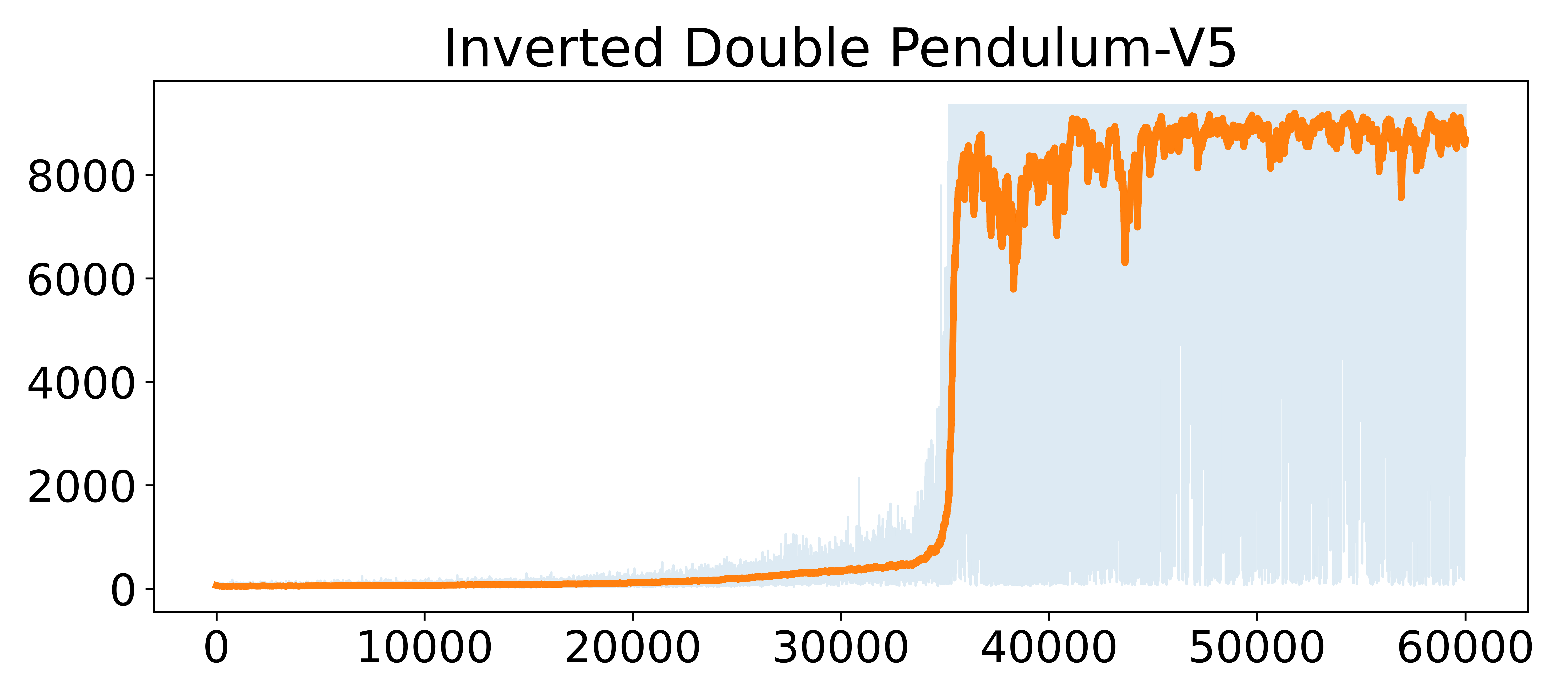}
       \vspace{-2mm}
       \panelcaption{(b) InvertedDoublePendulum-v5}
   \end{minipage}

   \vspace{2mm}

   \begin{minipage}{0.49\textwidth}
       \centering
       \includegraphics[width=\linewidth]{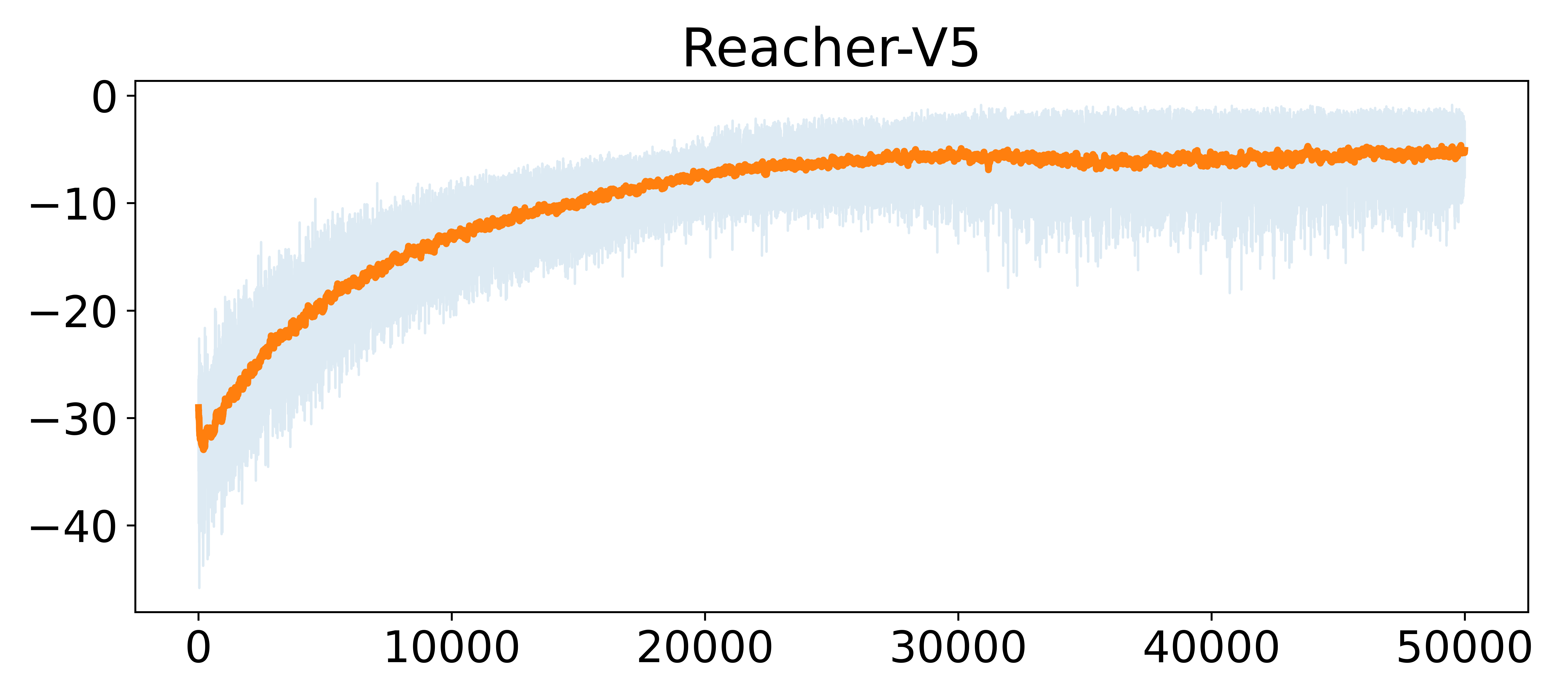}
       \vspace{-2mm}
       \panelcaption{(c) Reacher-v5}
   \end{minipage}
   \hfill
   \begin{minipage}{0.49\textwidth}
       \centering
       \includegraphics[width=\linewidth]{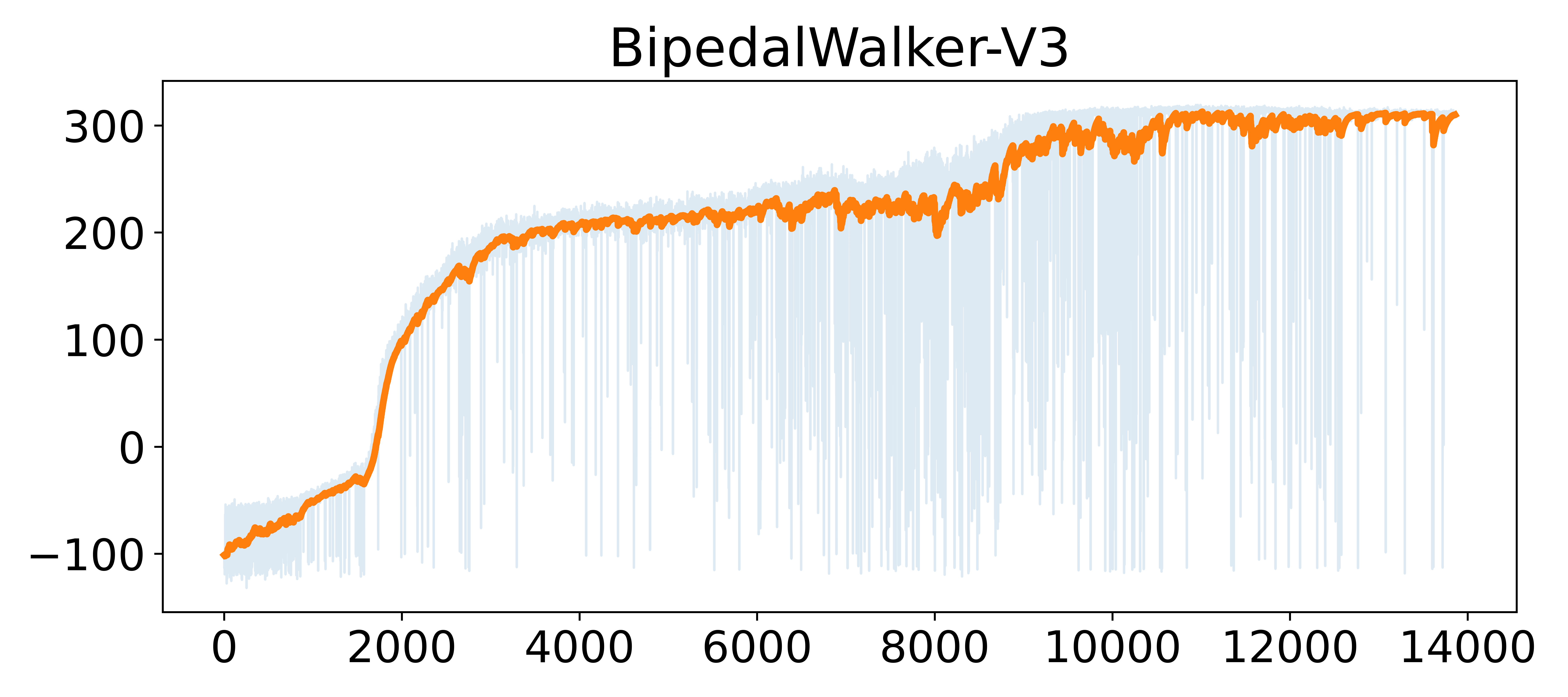}
       \vspace{-2mm}
       \panelcaption{(d) BipedalWalker-v3}
   \end{minipage}

   \vspace{2mm}

   \begin{minipage}{0.62\textwidth}
       \centering
       \includegraphics[width=\linewidth]{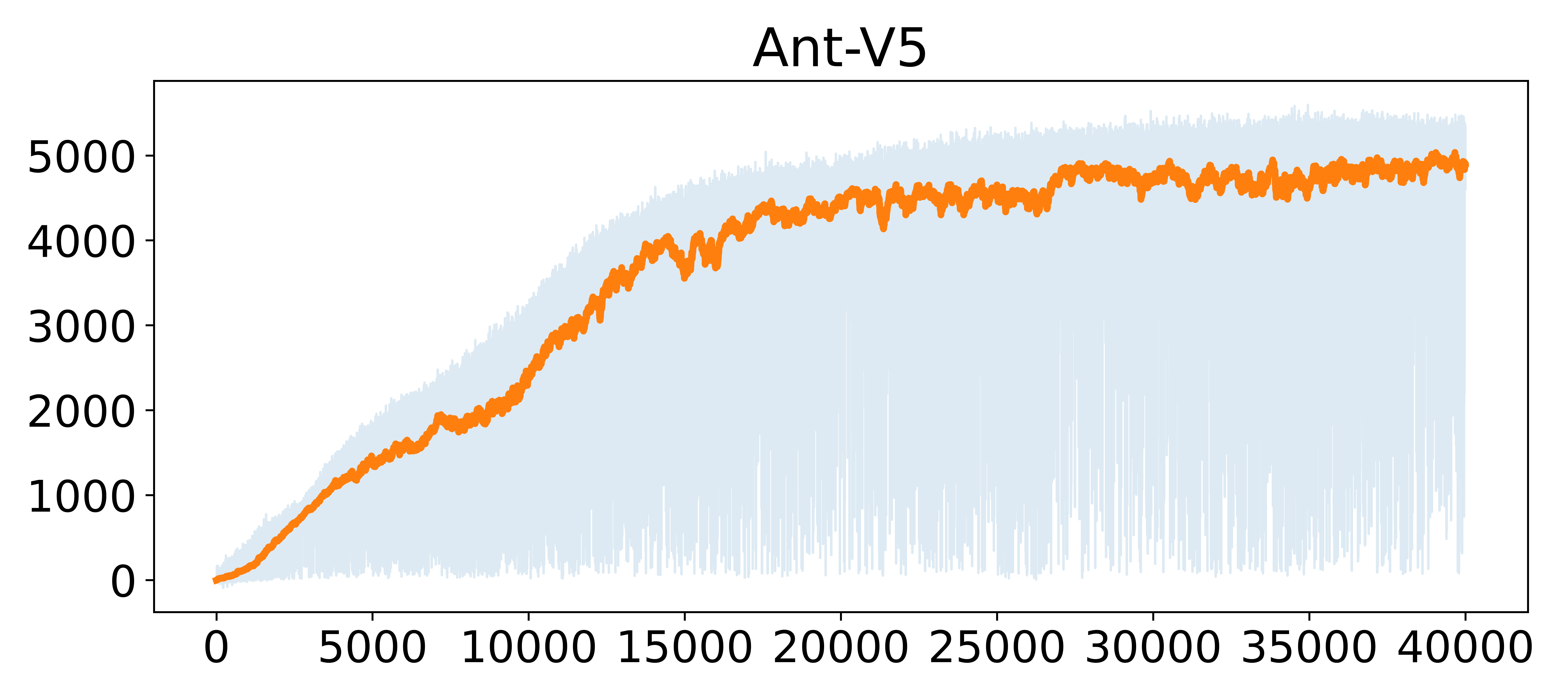}
       \vspace{-2mm}
       \panelcaption{(e) Ant-v5}
   \end{minipage}

   \caption{
   Additional SR-HOM training-return curves on five continuous-control benchmarks.
   Each panel shows the raw episodic return together with a smoothed return curve.
   The raw curve reflects the intrinsic stochasticity and episode-to-episode fluctuations during reinforcement-learning training, whereas the smoothed curve highlights the overall learning trend and final performance regime.
   }
   \label{fig:supp_srhom_return_curves}
\end{figure}

Fig.~\ref{fig:supp_srhom_return_curves} presents the training-return trajectories of the proposed SR-HOM agent across the five continuous-control benchmark environments. Each panel shows the raw episodic returns together with a smoothed curve, thereby capturing both episode-to-episode variability and the underlying learning trend. These trajectories complement the quantitative results reported in Table~\ref{tab:supp_metrics_all_envs} and provide a direct view of the learning dynamics of the SR-HOM agent across tasks involving low-dimensional control, balance, reaching, and locomotion.

For better visualization, an exponential moving average (EMA) is applied in the learning-curve figures. Given a raw sequence of episodic returns $\{R_t\}$, where $R_t$ denotes the return obtained in episode $t$, the smoothed curve is computed recursively as

\[
\bar{R}^{\mathrm{EMA}}_t=
\alpha R_t+(1-\alpha)\bar{R}^{\mathrm{EMA}}_{t-1},
\]
where $\alpha\in(0,1]$ determines the degree of smoothing. The EMA reduces high-frequency episode-to-episode fluctuations while retaining the overall learning trend.

Quantitative peak performance, by contrast, is evaluated using the maximum 100-episode simple moving-average return (SMA100),
\[
\mathrm{SMA100}_{\max}=
\max_t\left[
\frac{1}{100}\sum_{k=t}^{t+99}R_k
\right],
\]
where the maximum is taken over all valid 100-episode windows. The two averages therefore serve distinct purposes: the EMA provides a clearer visualization of the training dynamics, whereas $\mathrm{SMA100}_{\max}$ provides an interpretable quantitative metric that is independent of the EMA smoothing parameter and visualization procedure.

\section{Online Reinforcement-Learning Calibration of Tunable-Coupler Two-Qubit Gates}

This section first introduces tunable-coupler two-qubit gates implemented with superconducting transmon qubits and describes the calibration challenges caused by temporal drift in device and control parameters. We then formulate online gate calibration as a reinforcement-learning problem by specifying the simulation environment, the experimentally accessible observations available to the agent, the continuous control parameters that constitute the action space, and the reward used to quantify gate performance.

\subsection{Drifted tunable-coupler simulation}

We model a three-mode \(Q_1\)--\(C\)--\(Q_2\) module, where \(Q_1\) and
\(Q_2\) are fixed-frequency transmon qubits and \(C\) is a tunable coupler.
A time-dependent flux pulse applied to the coupler control line modulates the
coupler frequency and thereby activates and shapes the effective interaction
between the two qubits. The device parameters are chosen according to
Ref.~\cite{Li2025PulseCalibration}. Each mode is truncated to its three lowest
energy levels, allowing the simulation to capture coherent population transfer
out of the computational subspace and the resulting leakage.

For each applied waveform, we solve the time-dependent Schrödinger equation
and extract the effective operation projected onto the two-qubit computational
subspace. From this projected operation, we evaluate the leakage, conditional
phase, residual single-qubit phases, and gate fidelity. Energy relaxation and
dephasing are not included in the present model. The simulated degradation of
the gate therefore arises solely from coherent pulse distortion, parameter
drift, and leakage.

Setting $\hbar=1$, the Hamiltonian used to simulate the tunable coupler system is
\begin{align}
H(t)
=
\sum_{j\in\{1,c,2\}}
\left[
\omega_j(t)n_j
-\frac{\alpha_j}{2}n_j(n_j-1)
\right]
+
g_{1c}\left(a_1^\dagger a_c+a_1a_c^\dagger\right)
+
g_{2c}\left(a_2^\dagger a_c+a_2a_c^\dagger\right)
+
g_{12}\left(a_1^\dagger a_2+a_1a_2^\dagger\right).
\end{align}
Here, $a_j$ and $n_j=a_j^\dagger a_j$ denote the annihilation and number operators of mode $j$, respectively. The parameter $\omega_j(t)$ is the angular frequency of the $|0\rangle\leftrightarrow|1\rangle$ transition, $\alpha_j>0$ is the magnitude of the mode anharmonicity, and $g_{ij}$ is the transverse exchange-coupling strength between modes $i$ and $j$. The computational-qubit frequencies, $\omega_1$ and $\omega_2$, are held fixed throughout the simulation, whereas the coupler frequency $\omega_c(t)$ is modulated by the distorted flux waveform introduced below.

\paragraph{Flux-to-frequency model.}

The waveform arriving at the coupler is represented by a normalized flux coordinate $\Phi(t)$. In the reduced numerical model, this coordinate is converted into the instantaneous coupler transition frequency according to
\begin{equation}
\frac{\omega_c(t)}{2\pi}
=
f_{c,\mathrm{idle}}
-
\Delta f_c\,
\operatorname{clip}\!\left[\Phi(t),0,1.25\right],
\end{equation}
where $f_{c,\mathrm{idle}}=7.6635~\mathrm{GHz}$ is the coupler frequency at the idle operating point. The clipping operation restricts the effective flux coordinate to the interval $[0,1.25]$, while $\Delta f_c$ sets the frequency-excursion scale associated with the normalized control waveform.

Rather than being taken directly from a device-level flux-to-frequency relation, $\Delta f_c$ is determined by numerical calibration of the nominal clean pulse. For the CZ-gate configuration used here, we scan 251 uniformly spaced values of $\Delta f_c$ over an interval and select the value that maximizes the corresponding gate fidelity after applying the baseline virtual-$Z$ phase compensation.

\paragraph{Time-dependent propagation.}

Using a time step of $\Delta t=0.1~\mathrm{ns}$, we approximate the time-dependent evolution by treating the Hamiltonian as constant within each interval. The resulting propagator is evaluated as
\begin{align*}
U(T)=
U_{N-1}U_{N-2}\cdots U_1U_0,
\end{align*}
where $U_n=\exp\!\left[-iH\!\left(\omega_c(\Phi_n)\right)\Delta t\right]$, and $\Phi_n$ is the effective flux-control coordinate during the $n$th time interval, with $T=N\Delta t$.

The CZ and iSWAP control sequences consist of a gate segment, followed by a $6~\mathrm{ns}$ post-pulse segment and a $10~\mathrm{ns}$ idle segment. The gate and post-pulse segments are explicitly included in the quantum-state propagation. The idle segment is not included in the coherent gate evolution, but is retained in the applied command sequence because it contributes to the control-line memory carried into subsequent operations, following the pulse-sequence structure of Ref.~\cite{Li2025PulseCalibration}.

\paragraph{Control-line distortion model.}

The waveform produced by the pulse compiler is not assumed to reach the coupler without distortion. 
Let \(u_{k,n}=u_k(n\Delta t)\) be the commanded waveform at sample \(n\) in online window \(k\), with \(n=0,\ldots,N_k-1\). 
We model the hidden control-line state at the beginning of the window as
\[
\mathbf h_k
=
\left(
\delta\Phi_k,\,
\epsilon_{g,k},\,
p_{1,k},\,
p_{2,k},\,
m_{1,k}^{\rm in},\,
m_{2,k}^{\rm in}
\right).
\]
Here \(\delta\Phi_k\) is an additive flux-offset error, \(\epsilon_{g,k}\) is a fractional gain error, \(p_{\ell,k}\) is the strength of the \(\ell\)th transient-distortion mode, and \(m_{\ell,k}^{\rm in}\) is the residual memory of that mode at the beginning of the window. 
The corresponding gain factor is \(1+\epsilon_{g,k}\).

Within the window, each transient mode is modeled as an exponentially decaying memory driven by changes in the commanded waveform. 
We set \(m_{\ell,k,-1}=m_{\ell,k}^{\rm in}\) and define \(u_{k,-1}\) as the last command sample immediately before the window begins. 
The memory then evolves as
\begin{equation}
m_{\ell,k,n}
=
\rho_\ell m_{\ell,k,n-1}
+
p_{\ell,k}
\left(
u_{k,n}-u_{k,n-1}
\right),
\qquad
\rho_\ell
=
\exp\!\left(-\frac{\Delta t}{\tau_\ell}\right),
\label{eq_m}
\end{equation}
where \(\tau_\ell\) is the decay time of the \(\ell\)th mode. 
At the end of the window, the final memory value is carried forward, \(m_{\ell,k+1}^{\rm in}=m_{\ell,k,N_k-1}\), unless the simulation explicitly resets the line state. Specifically, We set \(m_{\ell,k,-1}=m_{\ell,k}^{\rm in}\), update \(m_{\ell,k,n}\) recursively within the window, and carry the final value \(m_{\ell,k,N_k-1}\) into the next window as \(m_{\ell,k+1}^{\rm in}\).

The effective dimensionless flux-control coordinate seen by the coupler is
\begin{equation}
\Phi_{k,n}
=
\delta\Phi_k
+
\left(1+\epsilon_{g,k}\right)
\left[
u_{k,n}
+
\sum_{\ell=1}^{2}m_{\ell,k,n}
\right].
\label{eq_phi}
\end{equation}
Thus, \(\delta\Phi_k\) shifts the waveform, \(\epsilon_{g,k}\) rescales it, and \(m_{\ell,k,n}\) introduces history-dependent transient distortion. 
The resulting coordinate \(\Phi_{k,n}\) is then converted into the instantaneous coupler frequency \(\omega_c(\Phi_{k,n})\).

\paragraph{Episode initialization and stochastic drift.}

At the beginning of each episode, the offset and fractional gain error are sampled as
\[
\delta\Phi_0
\sim
\mathcal U(-\delta_0,\delta_0),
\qquad
\epsilon_{g,0}
\sim
\mathcal U(-\epsilon_0,\epsilon_0),
\]
with $\delta_0=\epsilon_0=3\times10^{-3}$. For the nominal CZ configuration, the transient amplitudes and decay times are initialized at their reference values, while the residual memory is initialized according to the pulse history represented in the simulated command sequence.

Between consecutive online windows, the hidden parameters undergo independent stochastic updates,
\begin{align}
\delta\Phi_{k+1}
&=
\delta\Phi_k
+
\sigma_{\Phi}\xi_{\Phi,k}
+
J_k\sigma_J\xi_{J,k},
\nonumber\\
\epsilon_{g,k+1}
&=
\epsilon_{g,k}
+
\sigma_g\xi_{g,k},
\nonumber\\
p_{\ell,k+1}
&=
p_{\ell,k}
+
\sigma_p\xi_{\ell,k},
\qquad \ell\in\{1,2\},
\end{align}
where all $\xi$ variables are independent standard normal random variables. The Bernoulli variable $J_k$ models an occasional abrupt offset jump and satisfies
$J_k\sim\operatorname{Bernoulli}(q_J)$.
The numerical values used here are
$
\sigma_{\Phi}=1.2\times10^{-4}, 
\sigma_g=8.0\times10^{-5}, 
\sigma_p=2.5\times10^{-5}, 
q_J=0.005, 
\sigma_J=1.5\times10^{-3}$.

For the CZ simulation, the two nominal transient modes are taken from the measured response of the $C_{78}$ coupler reported in Ref.~\cite{Li2025PulseCalibration}:
\[
\begin{aligned}
\left(p_{1,0},p_{2,0}\right)
&=
(-0.019,-0.021),\\
\left(\tau_{1},\tau_{2}\right)
&=
(47.83,528.10)~\mathrm{ns}.
\end{aligned}
\label{eq:c78_transient_parameters}
\]
The decay times are fixed in the environment, whereas the offset, gain, and transient amplitudes evolve during each episode. These hidden quantities are not directly supplied to the agent; the controller must infer their effect from noisy gate-diagnostic measurements.

At the beginning of each episode, \(p_1\), \(p_2\), \(\tau_1\), and
\(\tau_2\) are initialized to these nominal values, while the control-line
gain and offset are independently varied within \(\pm 0.003\).
During the episode, the offset, gain, and transient amplitudes \(p_\ell\)
undergo independent Gaussian random walks. The time constants \(\tau_\ell\) remain fixed. To model occasional abrupt drift, an additional offset jump with standard deviation \(1.5\times10^{-3}\) occurs with probability \(0.005\) at each
step.

\subsection{Bounded incremental pulse and phase corrections}

This subsection defines the action space used by the SR-HOM agent and describes
how the policy output is converted into physically admissible corrections to
the coupler pulse and the virtual-\(Z\) phases. Rather than synthesizing an
unconstrained waveform at every online step, the agent applies bounded
incremental modifications to a previously calibrated reference control
sequence. This parameterization restricts the search to a compact neighborhood
of the nominal solution and reduces the risk of producing experimentally
unrealistic pulses.

At online step $k$, the action vector is defined as
\begin{equation}
\mathbf a_k
=
\left(
\Delta A_k,\,
\Delta b_k,\,
c_{{\rm fast},k},\,
c_{{\rm slow},k},\,
c_{{\rm post},k},\,
\Delta\phi_{Z1,k},\,
\Delta\phi_{Z2,k}
\right),\quad \mathbf a_k\in[-1,1]^7
\end{equation}

The first five components modify the analog flux-control pulse applied to the tunable coupler. The amplitude correction $\Delta A_k$ rescales the main pulse, while the bias correction $\Delta b_k$ introduces an additive offset. The coefficients $c_{{\rm fast},k}$ and $c_{{\rm slow},k}$ control corrective waveform components designed to compensate transient distortions on short and long timescales, respectively, and $c_{{\rm post},k}$ adjusts a short post-pulse segment used to mitigate residual control-line memory. The remaining two components, $\Delta\phi_{Z1,k}$ and $\Delta\phi_{Z2,k}$, specify virtual-$Z$ frame updates in software for the two computational qubits and compensate residual local phases without modifying the analog coupler waveform.

The actor produces a normalized action vector $\mathbf a_k\in[-1,1]^7,$ 
which is mapped to a bounded incremental update of the physical correction parameters. Let $\mathbf p_k$ denote the accumulated pulse and virtual-$Z$ corrections at online step $k$. The physical correction  parameter update is defined by
\begin{equation}
\mathbf p_{k+1}
=
\operatorname{clip}
\left(
\mathbf p_k
+
\mathbf s_a\odot\mathbf a_k,\,
-\mathbf p_{\max},\,
\mathbf p_{\max}
\right),
\label{eq:bounded_action_update}
\end{equation}

where $\mathbf s_a$ specifies the component-dependent update scale and $\mathbf p_{\max}$ sets the corresponding admissible bounds. The symbol $\odot$ denotes element-wise multiplication, and the clipping operation is applied independently to each component. This incremental and component-wise bounded parameterization limits the change applied at each online step, prevents unphysically large deviations from the calibrated reference pulse, and keeps the controller within a predefined admissible control region.

\paragraph{Reference pulse and analog corrections.}

The clean reference waveform used in the numerical environment is a tanh-shaped flat-top pulse
\[
u_0(t)
=
\frac{1}{2}
\left[
\tanh\!\left(\frac{t}{t_r}\right)
-
\tanh\!\left(\frac{t-T_g}{t_r}\right)
\right],
\]
defined over the main gate interval $0\leq t<T_g$, with CZ gate duration
$T_g=45~\mathrm{ns}$ and tanh-edge timescale $t_r=4~\mathrm{ns}$.

Given the accumulated correction vector $\mathbf p_k$ at online step $k$, the corrected waveform during the main gate segment is
\begin{align*}
u_{\mathrm{gate}}(t;\mathbf p_k)
={}
(1+\Delta A_k)u_0(t)
+\Delta b_k
+
c_{{\rm fast},k}
e^{-t/(50~\mathrm{ns})}u_0(t)
+
c_{{\rm slow},k}
e^{-t/(500~\mathrm{ns})}u_0(t).
\end{align*}
Here, $\Delta A_k$ and $\Delta b_k$ provide multiplicative amplitude and additive bias corrections, respectively. The coefficients $c_{{\rm fast},k}$ and $c_{{\rm slow},k}$ weight two exponentially decaying correction basis functions associated with short- and long-timescale control-line distortions. Their characteristic times of $50~\mathrm{ns}$ and $500~\mathrm{ns}$ are rounded values chosen to approximately match the two transient-response timescales measured for the coupler control line in Ref.~\cite{Li2025PulseCalibration}.

\paragraph{Post-pulse and idle segments.}

Following the main gate segment, a constant post-pulse correction of amplitude $c_{{\rm post},k}$ is applied for
$T_{\rm post}=6~\mathrm{ns}$. This is followed by a zero-command idle segment of duration
$T_{\rm idle}=10~\mathrm{ns}$. Defining $T_{\rm tot}
=
T_g+T_{\rm post}+T_{\rm idle}$,
the complete commanded waveform is
\begin{equation}
u(t;\mathbf p_k)
=
\begin{cases}
u_{\mathrm{gate}}(t;\mathbf p_k),
& 0\leq t<T_g,\\[2mm]
c_{{\rm post},k},
& T_g\leq t<T_g+T_{\rm post},\\[2mm]
0,
& T_g+T_{\rm post}\leq t<T_{\rm tot}.
\end{cases}
\end{equation}

The effective flux-control coordinate experienced by the coupler contains both the response to the current commanded waveform and residual transients generated by previously applied pulses. The residual response of the $\ell$th transient mode is updated according to Eq.~(\ref{eq_m}) and Eq.~(\ref{eq_phi}).

The memory variables are propagated continuously through the gate, post-pulse, and idle segments and are not reset between consecutive CZ operations. Consequently, the effective flux waveform applied during a given gate depends not only on its current command but also on the recent pulse history. This memory effect is particularly relevant for repeated-gate diagnostic sequences, in which residual transients can accumulate across successive CZ applications.

\paragraph{Virtual-\(Z\) corrections.}

The virtual-$Z$ corrections are implemented separately as software-defined frame updates on the two computational qubits. They compensate residual local phases without modifying either the commanded coupler waveform $u_{k,n}$ or the distorted flux coordinate $\Phi_{k,n}$.

The analog pulse corrections and virtual-\(Z\) updates together provide a
seven-dimensional action space that addresses both waveform distortion and
coherent phase error while remaining centered on the nominally calibrated
gate.

\subsection{Measurement-based observations}

This subsection describes the design of the environment observations provided to the agent during online calibration.

The policy receives diagnostic information that can, in principle, be obtained from low-overhead calibration circuits. At online step $k$, the observation includes the normalized accumulated correction parameters $\mathbf p_k\oslash\mathbf p_{\max}$, where $\oslash$ denotes elementwise division,
repeated gate-specific diagnostics, an estimated leakage probability, two residual local-phase estimates, the previous action, and the normalized episode progress $k/k_{\total}$. The gate-specific diagnostics are repeated-CZ conditional-phase errors for the CZ task and repeated-iSWAP swap-population errors together with a residual conditional-phase estimate for the iSWAP task.

Because direct estimation of the full gate fidelity is experimentally costly, the true fidelity is excluded from the observation and is not used explicitly in the training reward. It is computed only during validation for performance evaluation and reporting.

Leakage at online step $k$ is defined as the average probability that the four computational-basis input states evolve outside the computational subspace:
\[
L_k
=
1-
\frac{1}{4}
\sum_{x\in\{00,01,10,11\}}
\left\|
P_{\rm comp}U_k
\ket{x,0_c}
\right\|^2,
\]
where $U_k$ is the gate propagator generated by the corrected waveform at step $k$, and
\[
P_{\rm comp}
=
\sum_{q_1,q_2\in\{0,1\}}
\ket{q_1,0_c,q_2}
\bra{q_1,0_c,q_2}
\]
projects onto the two-qubit computational subspace with the coupler in its ground state. Thus, $L_k$ captures population transferred either to higher qubit levels or to excited coupler states.

To model finite-shot measurement noise, each leakage diagnostic is sampled from a binomial distribution,
\[
n_{{\rm leak},k}
\sim
\operatorname{Binomial}\!\left(N_L,L_k\right),
\qquad
\widehat{L}_k
=
\frac{n_{{\rm leak},k}}{N_L},
\]
where $n_{{\rm leak},k}$ is the number of detected leakage events and $N_L=512$ is the number of shots used for each leakage estimate. The resulting quantity $\widehat{L}_k$, rather than the exact simulated value $L_k$, is supplied to the agent.

The residual local phases, $\widehat{\phi}_{1,k}$ and $\widehat{\phi}_{2,k}$, are estimated using $N_Z=256$ measurement shots per phase diagnostic. Each conditional-phase estimate used in the CZ calibration task is likewise obtained from $N_\phi=256$ shots. These finite shot counts determine the statistical measurement noise present in the diagnostic observations supplied to the policy.

The remaining gate-specific diagnostics depend on the calibration target. For the CZ task, the policy receives conditional-phase errors extracted from repeated-CZ sequences. For the iSWAP task, the corresponding observation contains repeated-gate swap-population errors together with a diagnostic estimate of the residual conditional phase.

For the CZ calibration task, a flux pulse applied to the tunable coupler shifts its frequency from the idle operating point toward an interaction region. This activates an effective $ZZ$ interaction between the computational qubits. The resulting evolution is designed to accumulate a conditional phase of $\pi$ on the $\ket{11}$ state relative to the other computational-basis states, after removing the global and single-qubit phases. This mechanism corresponds to the coupler-only adiabatic CZ implementation considered in Ref.~\cite{Li2025PulseCalibration}.

Let
\[
U_{{\rm comp},k}
=
P_{\rm comp}U_k(T)P_{\rm comp}
\label{eq:projected_cz_propagator}
\]
denote the propagator generated by the corrected waveform at online step $k$, projected onto the computational subspace and expressed in the ordered basis
$\{\ket{00},\ket{01},\ket{10},\ket{11}\}$. Because population may leave this subspace, $U_{{\rm comp},k}$ is not necessarily unitary. The diagonal phase associated with each computational-basis state is defined as
\[
\theta_{x,k}
=
\arg
\left[
\bra{x,0_c}U_k(T)\ket{x,0_c}
\right],
\qquad
x\in\{00,01,10,11\}.
\label{eq:computational_basis_phases}
\]
From these phases, we extract the raw single-qubit phases
\begin{align*}
\phi_{1,k}^{\rm raw}
&=
\operatorname{wrap}_{(-\pi,\pi]}
\left(
\theta_{10,k}-\theta_{00,k}
\right),
\nonumber\\
\phi_{2,k}^{\rm raw}
&=
\operatorname{wrap}_{(-\pi,\pi]}
\left(
\theta_{01,k}-\theta_{00,k}
\right),
\end{align*}
and the nonlocal conditional phase
\[
\phi_{ZZ,k}
=
\operatorname{wrap}_{(-\pi,\pi]}
\left(
\theta_{11,k}
-\theta_{10,k}
-\theta_{01,k}
+\theta_{00,k}
\right).
\]
The target CZ operation therefore requires
$\phi_{ZZ,k}
=
\pi
\quad
(\mathrm{mod}\ 2\pi)$,
together with low leakage and small residual local phases after virtual-$Z$ compensation.

The projected CZ fidelity score used in the numerical environment is

\[
F_{{\rm CZ},k}
=
\frac{
\left|
\operatorname{Tr}
\left[
U_{\rm CZ}^{\dagger}
V_{Z,k}
U_{{\rm comp},k}
\right]
\right|^2
+
\operatorname{Tr}
\left[
U_{{\rm comp},k}^{\dagger}
U_{{\rm comp},k}
\right]
}{d(d+1)},
\]
where $d=4$ and 
$U_{\rm CZ}
=
\operatorname{diag}(1,1,1,-1)$
is the ideal CZ operation and $V_{Z,k}$ represents the virtual-$Z$ frame compensation applied to remove the local single-qubit phases. This fidelity is computed from the simulated propagator only for validation and reporting and is not supplied directly to the policy.

The principal online phase diagnostic consists of repeated applications of the current CZ gate, with repetition numbers
$M\in\{1,2,4\}$.

For each value of $M$, one qubit is prepared on the equator of the Bloch sphere and used as the phase-sensitive target, while the other qubit is prepared in either $\ket{0}$ or $\ket{1}$. Let
$\widehat{\phi}^{(0)}_{M,k}$ and
$\widehat{\phi}^{(1)}_{M,k}$ denote the measured target-qubit phases for these two control-qubit preparations. Their difference provides an estimate of the accumulated conditional phase,
\[
\widehat{\phi}_{M,k}
=
\operatorname{wrap}_{(-\pi,\pi]}
\left(
\widehat{\phi}^{(1)}_{M,k}
-
\widehat{\phi}^{(0)}_{M,k}
\right),
\]
which is compared with the ideal value $M\pi$. The corresponding diagnostic error is
\[
\widehat{\delta\phi}_{M,k}
=
\operatorname{wrap}_{(-\pi,\pi]}
\left(
\widehat{\phi}_{M,k}
-
M\pi
\right).
\]

During each repeated-gate sequence, both the three-mode quantum state and the residual flux-line memory are propagated continuously without being reset between successive CZ applications. Consequently, coherent phase errors, leakage, and pulse-history-dependent distortions can accumulate across repetitions. Using several values of $M$ amplifies small systematic phase deviations and therefore improves the sensitivity of the diagnostic signal to control-parameter drift.

The iSWAP simulation only differs from the CZ simulation in target interaction. Whereas the CZ gate is generated by the accumulation of a conditional phase, the iSWAP gate is produced by coherent excitation exchange between the computational states $\ket{01}$ and $\ket{10}$. The ideal operation corresponds to a swap angle of $\pi/2$, together with minimal leakage, residual conditional phase, and local-phase error.

In the iSWAP configuration, the two computational qubits are tuned into resonance, and the coupler pulse activates the effective transverse exchange interaction for a fixed gate duration. The clean reference waveform retains the smooth-box form introduced for the CZ task, but uses a shorter gate duration $T_g=24~\mathrm{ns}$,
an edge-smoothing time of
$t_r=3~\mathrm{ns}$,
and post-pulse and idle durations of $T_{\rm post}=T_{\rm idle}=4~\mathrm{ns}$.

The principal iSWAP diagnostic probes the coherent population exchange generated by repeated applications of the current gate. At online step $k$, the corrected iSWAP operation is also applied $M\in\{1,2,3,4\}$ times. The swap probability is estimated in both exchange directions and averaged according to
\[
\widehat{P}_{{\rm swap},k}^{(M)}
=
\frac{1}{2}
\left[
\widehat{P}_{10\rightarrow 01,k}^{(M)}
+
\widehat{P}_{01\rightarrow 10,k}^{(M)}
\right],
\]
where $\widehat{P}_{10\rightarrow 01,k}^{(M)}$ and
$\widehat{P}_{01\rightarrow 10,k}^{(M)}$ are the measured transition probabilities after $M$ repeated gates for the two exchange directions. For an ideal iSWAP operation with swap angle $\pi/2$, the expected population-transfer probability is
\[
P_{\rm swap,ideal}^{(M)}
=
\sin^2\!\left(\frac{M\pi}{2}\right).
\]
The corresponding measured swap-population error is therefore defined as
\[
\widehat{e}_{M,k}^{\rm swap}
=
\widehat{P}_{{\rm swap},k}^{(M)}
-
\sin^2\!\left(\frac{M\pi}{2}\right).
\]

In addition to the repeated-gate swap errors, the diagnostic observation includes an estimate of the residual conditional phase,
$\widehat{\phi}_{ZZ,k}^{\rm res}$, the leakage estimate
$\widehat{L}_k$, and the two residual local-phase estimates
$\widehat{\phi}_{1,k}$ and $\widehat{\phi}_{2,k}$. The same finite-shot allocations used for the corresponding CZ diagnostics are adopted in the iSWAP simulation.

\subsection{Reward function}

This section describes the design of the reward function used to train the agent for online gate calibration.

The CZ reward penalizes deviations from the target conditional phase, leakage from the computational subspace, residual single-qubit phases, and excessively large or rapidly varying control actions. At online step $k$, $\delta\widehat{\phi}_{M,k}$ denotes the finite-shot estimate of the conditional-phase error accumulated after $M$ repeated CZ gates relative to the ideal value $M\pi$, while $\widehat{\phi}_{i,k}$ denotes the estimated residual local phase of qubit $i$. The reward is defined as
\begin{align*}
r_k
=
-\frac{w_\phi}{3}
\sum_{M\in\{1,2,4\}}
\left(\delta\widehat{\phi}_{M,k}\right)^2
-w_L\widehat{L}_k
-\frac{w_Z}{2}
\sum_{i=1}^{2}
\widehat{\phi}_{i,k}^{\,2}
\nonumber
-
\frac{w_a}{7}
\left\|\mathbf a_k\right\|_2^2
-
\frac{w_{\Delta a}}{7}
\left\|\mathbf a_k-\mathbf a_{k-1}\right\|_2^2 .
\end{align*}
Here, $\mathbf a_{k-1}=0$ $w_\phi$, $w_L$, and $w_Z$ control the penalties associated with conditional-phase error, leakage, and residual local phases, respectively. The coefficient $w_a$ regularizes the action magnitude, while $w_{\Delta a}$ suppresses abrupt changes between consecutive actions. The factors $1/3$, $1/2$, and $1/7$ normalize the corresponding sums by the number of repeated-gate diagnostics, computational qubits, and action components.

For the iSWAP task, the repeated-CZ conditional-phase penalty is replaced by a gate-specific term that penalizes both population-transfer errors and the residual conditional phase. The reward is defined as
\begin{align*}
r_k
=
-\frac{w_{\rm sp}}{5}
\left[
\sum_{M\in\{1,2,3,4\}}
\left(
\widehat{e}_{M,k}^{\rm swap}
\right)^2
+
\left(
\widehat{\phi}_{ZZ,k}^{\rm res}
\right)^2
\right]
-w_L\widehat{L}_k
-\frac{w_Z}{2}
\sum_{i=1}^{2}
\widehat{\phi}_{i,k}^{\,2}
\nonumber
-
\frac{w_a}{7}
\left\|\mathbf a_k\right\|_2^2
-
\frac{w_{\Delta a}}{7}
\left\|\mathbf a_k-\mathbf a_{k-1}\right\|_2^2 .
\label{eq:iswap_reward}
\end{align*}
Here, $\widehat{e}_{M,k}^{\rm swap}$ is the finite-shot estimate of the population-transfer error after $M$ repeated iSWAP gates, and $\widehat{\phi}_{ZZ,k}^{\rm res}$ is the estimated residual conditional phase. The first term therefore encourages the gate to reproduce the ideal repeated-iSWAP population dynamics while suppressing an unwanted $ZZ$ phase. The factor $1/5$ normalizes this contribution over the four repeated-gate swap diagnostics and the single residual-phase diagnostic. The remaining terms have the same roles as in the CZ reward.

\subsection{Online calibration workflow}

We begin with a reference pulse $u_0(t)$ obtained from a prior calibration. At the start of each episode, the environment samples a hidden, drifted control-line state under which the unmodified reference pulse no longer implements the target gate accurately. During online calibration window $k$, diagnostic measurements provide noisy estimates of the relevant gate errors, which are supplied to the SR-HOM agent as observations. The agent then outputs a bounded update to the pulse-correction parameters, and the resulting corrected waveform is applied in the subsequent window. This measurement--update cycle is repeated for a fixed number of windows, while the underlying drift state continues to evolve slowly throughout the episode.

\subsection{Training and evaluation}

\begin{figure*}[t]
    \centering

    \includegraphics[width=0.85\textwidth]{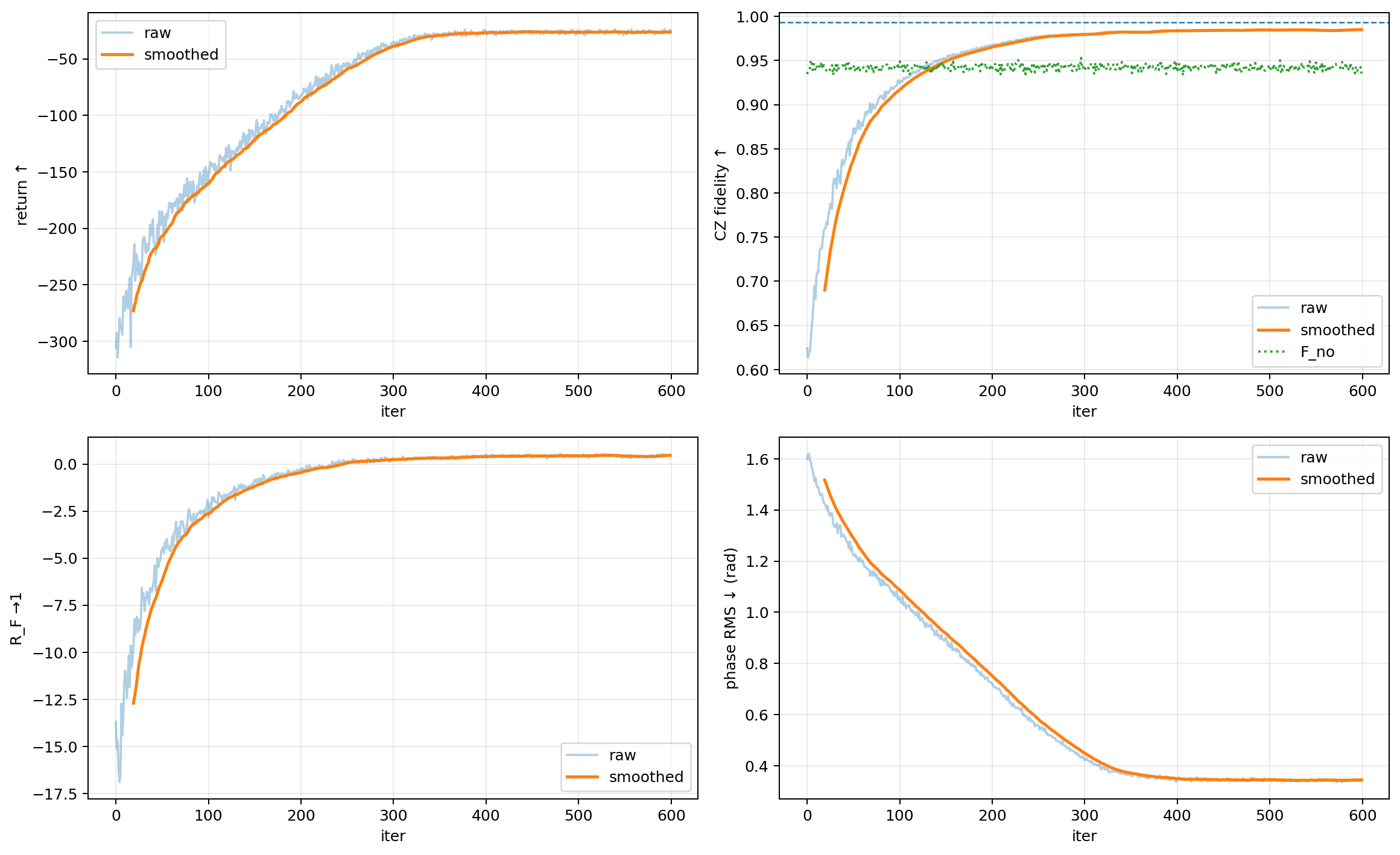}
    \vspace{0.3em}

    \textbf{(a)} CZ

    \vspace{1.0em}

    \includegraphics[width=0.85\textwidth]{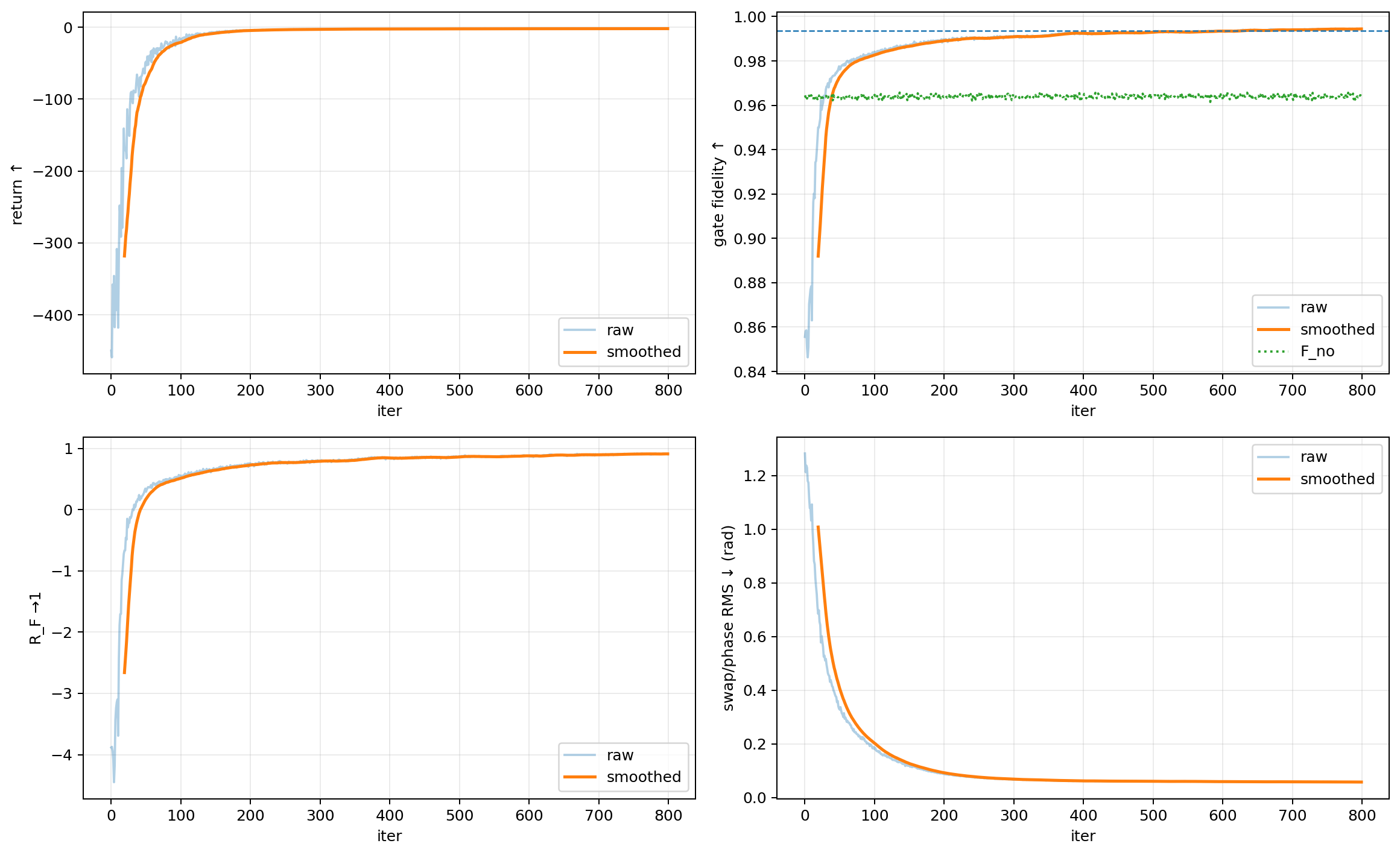}
    \vspace{0.3em}

    \textbf{(b)} iSWAP

    \caption{
    Training curves for measurement-based RL calibration of tunable-coupler two-qubit gates.
    (a) CZ calibration. (b) iSWAP calibration.
    In both tasks, the return improves during training, the RL-corrected fidelity rises above the no-correction baseline,
    and the diagnostic error is suppressed. The no-correction fidelity and clean-pulse reference are shown as horizontal baselines. The symbol $\to 1$ indicates that values closer to unity are preferable, while the upward and downward arrows denote metrics for which higher and lower values are better, respectively.
    }
    \label{fig:training_dynamics}
\end{figure*}

The training dynamics are shown in Fig.~\ref{fig:training_dynamics}. For both CZ and iSWAP, the episode return
increases during training and then approaches a plateau. At the same time, the corrected gate fidelity rises above the
no-correction baseline and approaches the clean-pulse reference, while the relevant diagnostic-error RMS decreases.
This behavior is notable because the policy is not optimized using the true fidelity. Instead, it learns to improve the
gate by reducing the measurement-based errors included in the observation and reward. The gate fidelity is evaluated separately to assess the physical performance of the learned calibration policy.

Let $F_{\rm clean}$ denote the gate fidelity corresponding to the clean reference pulse. When the same pulse
is applied after drift, the resulting no-correction fidelity is $F_{\rm no}$. After the policy updates the pulse
parameters, the corrected gate has fidelity $F_{\rm RL}$. The objective is to restore $F_{\rm RL}$ as close as possible
to $F_{\rm clean}$ and limit leakage. We quantify the recovered fraction of the drift-induced fidelity loss by
\[
R_F=\frac{F_{\rm RL}-F_{\rm no}}{F_{\rm clean}-F_{\rm no}}.
\]
Here, $R_F < 0$ means the RL agent worsens the fidelity, $R_F\leq 0$ means that the controller provides no improvement over the drifted pulse, whereas $R_F\geq 1$
means that it fully recovers the clean calibrated reference and may improve the pulse beyond that reference. By the time of the first measurement, the system has already accumulated some drift and continues to drift slowly during the calibration process. And we define the relative fidelity with respect to the clean-pulse reference as
\[
\eta_F
=
\frac{F_{\rm RL}}{F_{\rm clean}}.
\]
Thus, $\eta_F=1$ indicates that the corrected gate
matches the clean-pulse reference, while $\eta_F<1$ indicates incomplete
recovery. Values $\eta_F>1$ are possible if the corrected pulse slightly outperforms the chosen clean reference.

\begin{figure*}[t]
    \centering

    \includegraphics[width=0.85\textwidth]{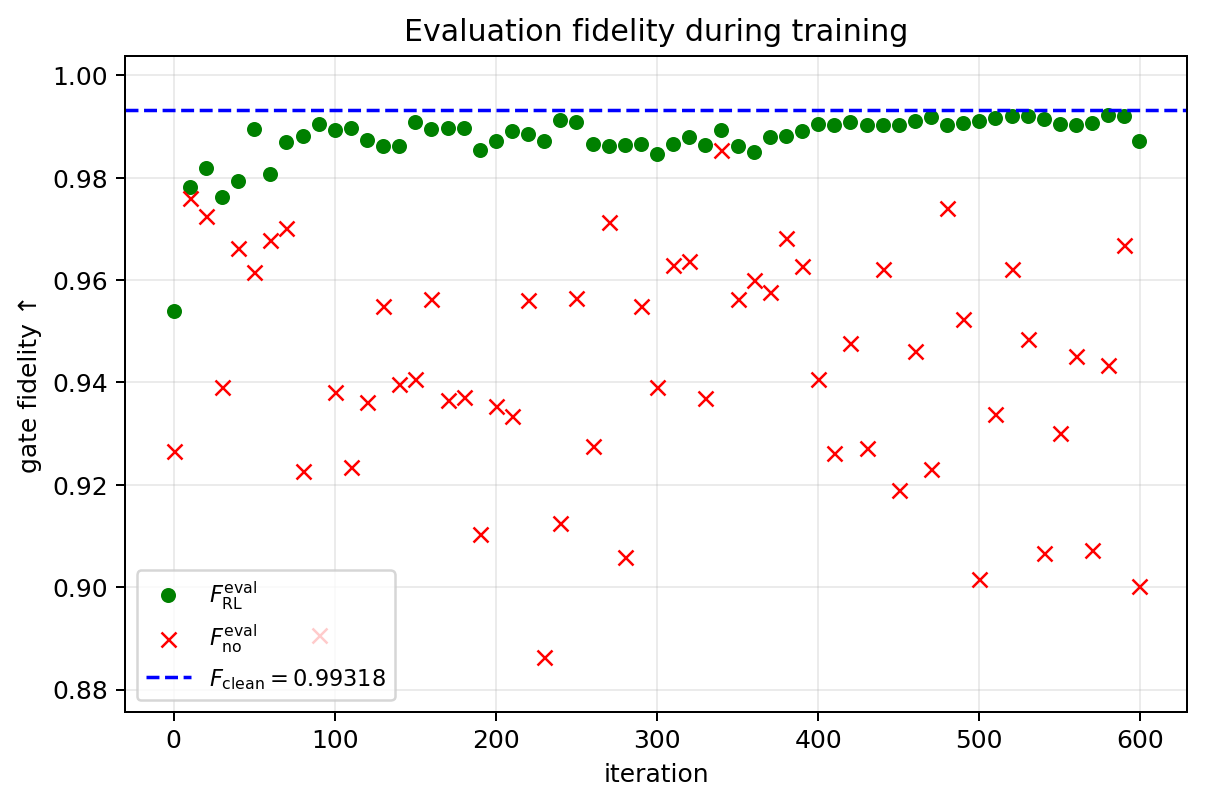}
    \vspace{0.2em}

    \textbf{(a)} CZ

    \vspace{1.0em}

    \includegraphics[width=0.85\textwidth]{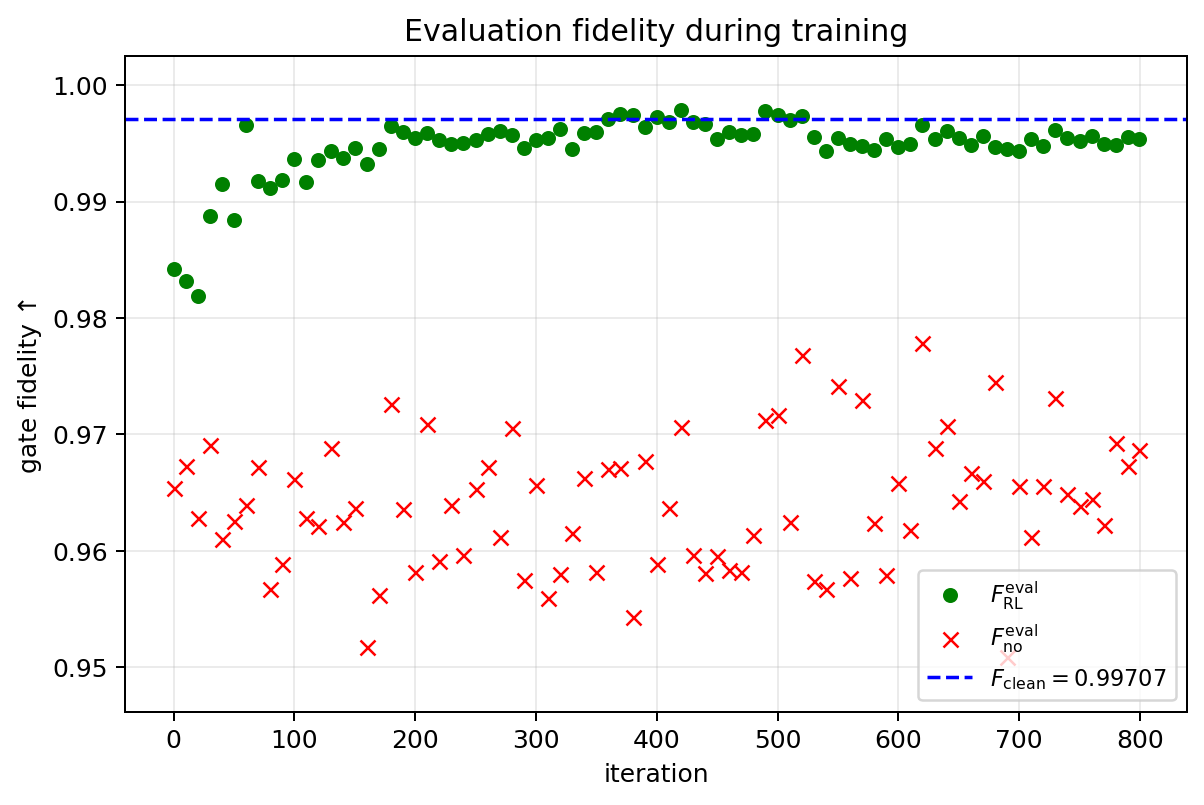}
    \vspace{0.2em}

    \textbf{(b)} iSWAP

    \caption{
    Evaluation fidelity during training for the CZ and iSWAP calibration tasks.
    The RL-corrected fidelity is compared with the paired no-correction fidelity and the clean-pulse reference.
    The learned policies rapidly improve the drifted pulses and maintain fidelities close to the clean references. The upward and downward arrows denote metrics for which higher and lower values are better, respectively.
    }
    \label{fig:eval_fidelity_training}
\end{figure*}

In addition to the quantities recorded during policy optimization, we periodically evaluate the current policies on
previously unseen drift conditions. The resulting fidelities are shown in Fig.~\ref{fig:eval_fidelity_training}. After
the initial learning period, the corrected fidelities remain above the corresponding no-correction baselines and close
to the clean references for both gates. The periodic evaluation therefore tests whether the policies learned during
training transfer to drift realizations not used for the policy update.

\begin{figure*}[t]
    \centering

    \includegraphics[width=0.80\textwidth]{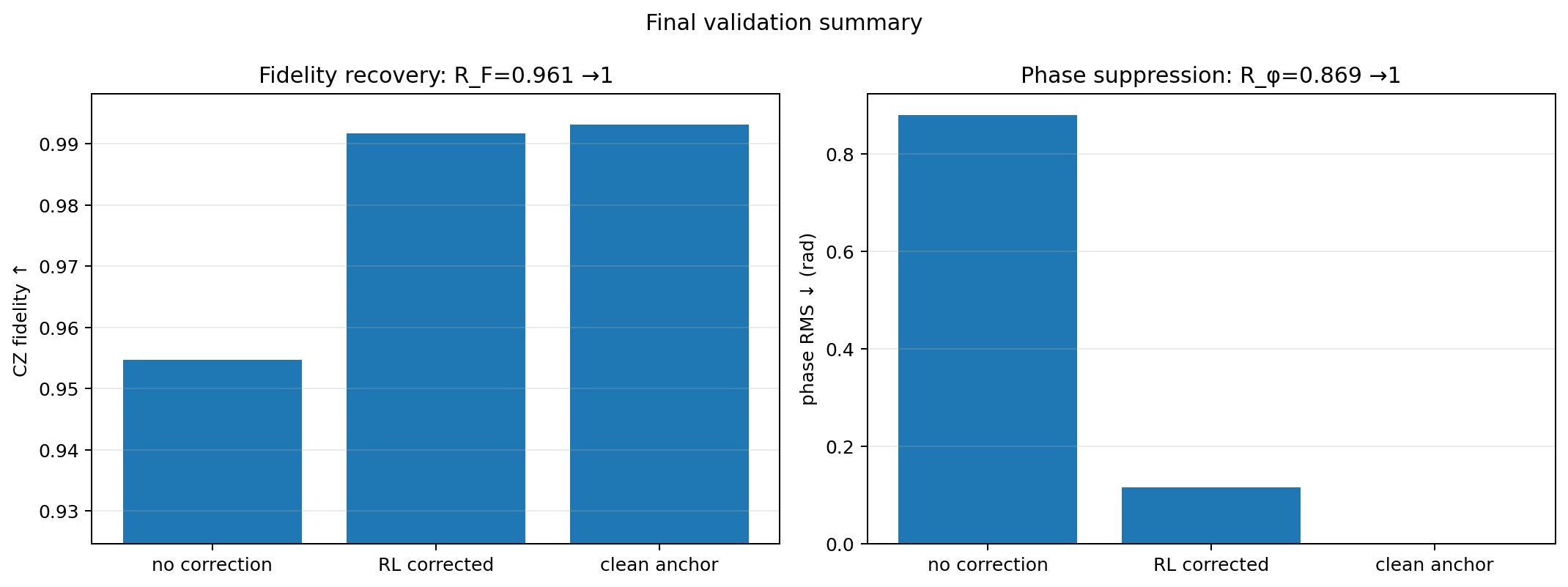}
    \vspace{0.2em}

    \textbf{(a)} CZ

    \vspace{1.0em}

    \includegraphics[width=0.80\textwidth]{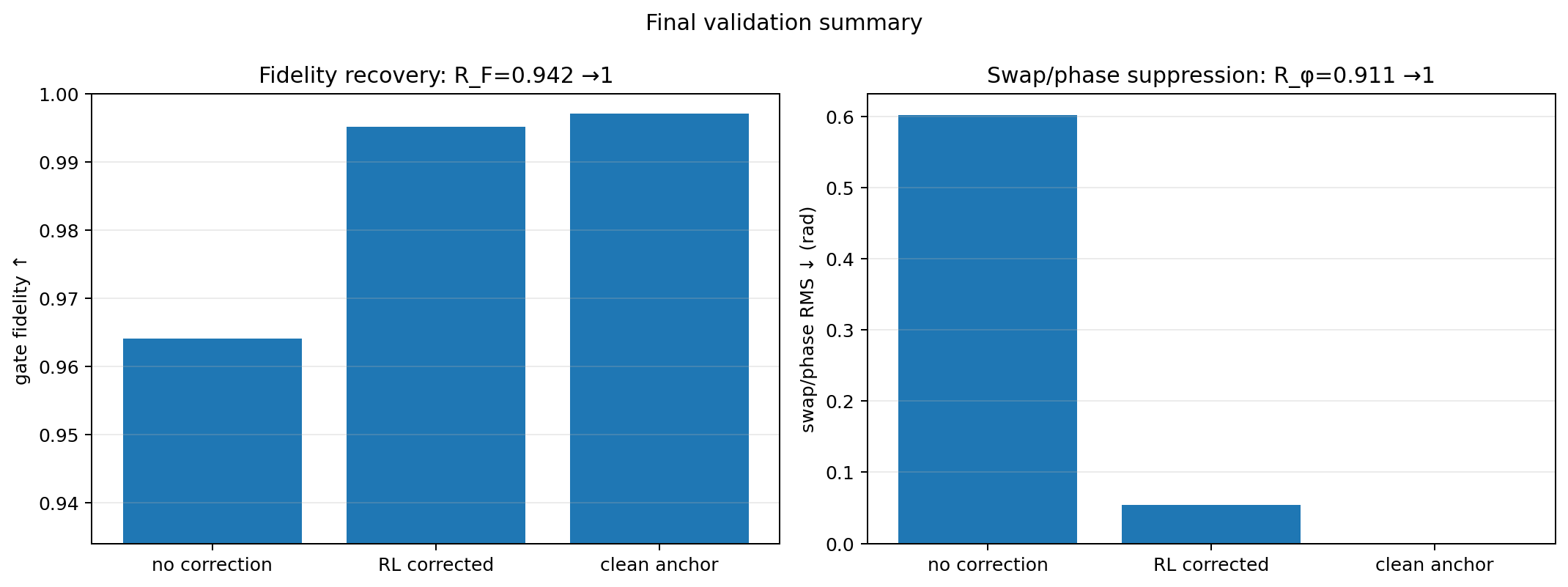}
    \vspace{0.2em}

    \textbf{(b)} iSWAP

    \caption{
    Final validation under previously unseen drift conditions. The symbol $\to 1$ indicates that values closer to unity are preferable.
    For both gates, the RL-corrected pulse restores the fidelity close to the clean reference while suppressing the repeated
    gate-specific error.
    }
    \label{fig:final_validation}
\end{figure*}

The final policies are evaluated on additional instances, with the results summarized in Fig.~\ref{fig:final_validation}. For the CZ task, the no-correction fidelity is $F_{\rm no}=0.9547$. The RL-corrected pulse reaches $F_{\rm RL}=0.9917$, close to the clean-pulse reference $F_{\rm clean}=0.9932$, recovering to about $\eta_F = 99.85\%$ of the reference fidelity. The corresponding recovery ratio is $R_F\simeq0.9610$, and the gate fidelity loss is reduced by approximately $81.6\%$. The repeated-CZ phase RMS decreases from $0.8802$ to $0.1154$, indicating that the fidelity recovery is primarily associated with suppression of the conditional-phase error. The mean leakage is reduced by approximately $27.39\%$, from $0.01481$ to $0.01075$.

For the iSWAP task, the independently evaluated no-correction fidelity is
$F_{\rm no}=0.9640$, whereas the RL-corrected pulse reaches
$F_{\rm RL}=0.9952$. The clean-pulse reference is
$F_{\rm clean}=0.9971$. The ratio of the mean fidelities gives
$R_F=0.9426$, while the mean of the fidelity recovery ratios is
$\eta_F = 99.81\%$. The gate fidelity loss is reduced by approximately
$86.6\%$ relative to the no-correction baseline. The mean finite-shot swap/phase RMS decreases from
$0.6016$ to $0.0536$, corresponding to a reduction of approximately
$91.1\%$. The mean finite-shot leakage estimate decreases from
$L_{\rm no}=0.0048$ to $L_{\rm RL}=0.0031$, which is a relative reduction of approximately
$35.4\%$.

Taken together, the CZ and iSWAP results show that the same measurement-based continuous-action controller can
recover most of the drift-induced fidelity loss for two distinct tunable-coupler gate mechanisms.

\end{document}